\documentclass[pra,twocolumn,preprintnumbers,amsmath,amssymb]{revtex4}

\usepackage{amscd}
\usepackage{graphicx}
\usepackage{dcolumn}
\usepackage{bm}

\newcommand{\proj}{\operatorname{proj}}  

\newcommand{\vect}[1] {\ensuremath{\mathbf{#1}}} 

\newcommand{\set}[1] {\ensuremath{\mathcal{#1}}} 
\newcommand{\mment}[1] {\ensuremath{\mathbf{#1}}} 

\newcommand{\numberfield}[1]{\ensuremath{\mathbb{#1}}} 

\newcommand{\cvect}[1] {\textsf{#1}} 

\newcommand{\lvect}[1] {\ensuremath{\mathbf{#1}}} 

\newcommand{\cmatrix}[1]{\textsf{#1}} 

\newcommand{\rmatrix}[1]{\ensuremath{#1}} 

\newcommand{\model}[1]{\ensuremath{\mathbf{#1}}} 

\newcommand{\ev}[1]{\ensuremath{\langle #1 \rangle}} 
\newcommand{\evb}[1]{\ensuremath{\bigl\langle #1 \bigr\rangle}} 

\newcommand{\var}{\ensuremath{\chi}} 

\addtocounter{secnumdepth}{1} 

\begin{document}


\title{An Information-Theoretic Approach to Quantum Theory,~II: \\
        The Formal Rules of Quantum Theory}

\author{Philip Goyal}
    \email{pg247@cam.ac.uk}
    \affiliation{Cavendish Laboratory \\
                University of Cambridge}


\begin{abstract}
In a companion paper~\cite{Goyal-QT1}~(hereafter referred to as
Paper~I), we have presented an attempt to derive the
finite-dimensional abstract quantum formalism from a set of
physically comprehensible assumptions.  In this paper, we formulate
a correspondence principle, the \emph{Average-Value Correspondence
Principle}, that allows relations between measurement outcomes which
are known to hold in a classical model of a system to be
systematically taken over into the quantum model of the system.
Using this principle, we derive the explicit form of the temporal
evolution operator~(thereby completing the derivation of the
abstract quantum formalism begun in Paper~I), and derive many of the
formal rules~(such as operator rules, commutation relations, and
Dirac's Poisson bracket rule) that are needed to apply the abstract
quantum formalism to model particular physical systems.

\end{abstract}

\maketitle

\section{Introduction}


In order to obtain a quantum mechanical model for a particular
physical system such as a particle moving in space, it is necessary
to supplement the abstract quantum formalism with formal rules which
explicitly determine the form of the operators that represent
particular measurements performed on the system, or that represent
particular symmetry transformations~(such as displacement or
rotation) of the frame of reference in which the system is being
observed.  These formal rules usually suppose that measurements are
described in the framework of classical physics, so that one speaks
of ``a measurement of observable~(or property)~$A$'' and of an
operator that represents such a measurement. These formal rules can
be usefully classified as follows:
\begin{itemize}
\item[(i)]\emph{Operator Rules.} Rules for writing down operators
representing measurements of observables that are known functions of
other, elementary, observables whose operators
are given~%
\footnote{The basic operator rules of quantum theory
are~(i)~Rule~1~(Function rule):~If a measurement of~$A$ is
represented by operator~$\cmatrix{A}$, then a measurement of~$f(A)$
is represented by~$f(\cmatrix{A})$; (ii)~Rule~2~(Sum rule):~If
measurements of~$A$ and~$B$ are represented by
operators~$\cmatrix{A}, \cmatrix{B}$, respectively, then a
measurement of~$f_1(A) + f_2(B)$ is represented by~$f_1(\cmatrix{A})
+ f_2(\cmatrix{B})$~(and similarly for more than two observables),
and (iii)~Rule~3~(Product rule):~If measurements of~$A$ and~$B$ are
represented by commuting operators~$\cmatrix{A}, \cmatrix{B}$,
respectively, then a measurement of~$f_1(A)f_2(B)$ is represented
by~$f_1(\cmatrix{A})f_2(\cmatrix{B})$~(and similarly for more than
two observables).  A more general rule that is often employed
is:~(iv)~Rule~3$'$~(Hermitization rule):~If measurements of~$A$
and~$B$ are represented by operators~$\cmatrix{A}, \cmatrix{B}$,
respectively, then a measurement of~$f_1(A)f_2(B)$ is represented
by~$(f_1(\cmatrix{A})f_2(\cmatrix{B}) +
f_2(\cmatrix{B})f_1(\cmatrix{A}))/2$.  The above rules are generally
assumed to apply to arbitrary Hermitian operators.}.
For example, such rules are needed to be able to write down the
operator that represents a measurement of~$H$, given the classical
relation~$H = p_x^2/2m + V(x)$, in terms of the
operators~$\cmatrix{x}$ and~$\cmatrix{p}_x$ which represent
measurements of~$x$ and~$p_x$, respectively.

\item[(ii)]\emph{Commutation Relations.}  Commutation relations
between operators that represent measurements of fundamental
observables such as position, momentum, and components of angular
momentum.  The commutation
relations~$[\cmatrix{x},\cmatrix{p}_x]=i\hbar$ and~$[\cmatrix{L}_x,
\cmatrix{L}_y]=i\hbar\cmatrix{L}_z$, are the obvious examples, while
Dirac's Poisson bracket rule,~$[\cmatrix{A}, \cmatrix{B}] = i\hbar
\widehat{\{A, B\}}$, is the more general rule for evaluating
commutation relations, where~$\{A,B\}$ is the classical Poisson
bracket for observables~$A$ and~$B$ of a physical system,
and~$\widehat{\{A,B\}}$, $\cmatrix{A}$, and~$\cmatrix{B}$ are the
respective operators.

\item[(iii)]\emph{Transformation Operators.}  Explicit forms of
the operators that represent symmetry transformations of a frame
of reference, such as the $x$-displacement operator~$\cmatrix{D}_x
= -i\, d/dx$.

\item[(iv)]\emph{Measurement--Transformation Relations.} The
relations between measurement operators and transformation
operators. For example, the $x$-displacement
operator,~$\cmatrix{D}_x$, stands in the relation~$\cmatrix{D}_x =
\cmatrix{p}_x/\hbar$ to the $x$-momentum measurement
operator,~$\cmatrix{p}_x$.
\end{itemize}

The physical origin of many of the above-mentioned rules is obscure.
For example, although one can give simple physical
arguments~\footnote{See~\cite{Dirac58}~(Sec.~11)
and~\cite{vNeumann55}~(Sec.~IV.1), for example.} for the operator
rule which states that, if a measurement of~$A$ is represented by
operator~$\cmatrix{A}$, then a measurement of a function,~$f(A)$,
of~$A$ is represented by operator~$f(\cmatrix{A})$, the
generalization of such arguments to measurements of functions of two
or more observables encounters severe difficulties due to the
possible non-commutativity of the operators that represent these
observables.  As a result of such difficulties, operator rules tend
to be heuristic and tend to lead to inconsistencies when applied to
particular examples~\footnote{See, for
example,~\cite{vNeumann55}~(Sec.~IV.1),
\cite{Bohm51}~(Sec.~9.12---9.15), and~\cite{Isham-QT}~(Sec.~5.2.1).
We shall discuss one such example in
Sec.~\ref{sec:inconsistencies}.}. Similarly, the commutation
relationship~$[\cmatrix{x},\cmatrix{p}_x]=i\hbar$ is typically
obtained from Schroedinger's equation or from Dirac's Poisson
bracket rule, both of whose derivations involve abstract assumptions
whose physical origin is obscure.

Recent work on the elucidation of the physical origin of the quantum
formalism either focuses exclusively on the derivation of the
abstract quantum formalism~\cite{Wootters80, Rovelli96,
Popescu-Rohrlich97, Caticha98b, Caticha99b, Summhammer99, Brukner99,
Brukner02a, Brukner02b, Clifton-Bub-Halvorson03, Grinbaum03,
Grinbaum04} or is concerned with the derivation of the Schroedinger
equation directly from informational ideas without taking the
abstract quantum formalism as a
given~\cite{Frieden-Schroedinger-derivation,
MJWHall-Schroedinger-derivation, Reginatto-Schroedinger-derivation}.
Consequently, the question of what additional physical ideas are
needed to obtain the formal rules described above \emph{given} the
abstract quantum formalism has received relatively little recent
attention.

Operator rules have been discussed in a few publications, for
example in~\cite{vNeumann55, Groenewold-Principles-QM}, but a
derivation of the operator rules on the basis of a physical
principle, taking the abstract formalism as a given, has not been
successfully completed. The most recent systematic attempt to
derive the commonly employed commutation relations and
measurement--transformation relations in such a manner is found
in~\cite{TFJordan75}~(see also Refs.~\cite{TFJordan69,
Ballentine98}). In~\cite{TFJordan75}, the commutation
relationships for the operators that represent the Galilean group
of transformations are first derived by exploiting the group
structure of the classically-described transformations.  Then, by
establishing the relation between these transformation operators
and particular measurement operators, commutation relations for
these measurement operators are obtained. However, this approach
implicitly makes use of operator rules, and relies upon auxiliary
assumptions, such as the assumption that certain measurement
operators are unchanged by the action of particular symmetry
operators, whose physical origin is unclear.

In this paper, we show that, starting from the abstract quantum
formalism, it is possible to derive the above-mentioned formal rules
in a straightforward manner from a correspondence principle. Roughly
speaking, this principle, the \emph{Average-Value Correspondence
Principle}~(AVCP), asserts that, in a classical experiment, if a
relation holds between the outcomes of a set of measurements
performed on a physical system, then the same relation holds
\emph{on average} in a corresponding, suitably-defined quantum
experiment on the same physical system, the average being taken over
infinitely many trials of the quantum experiment.

This paper is organized as follows. We begin in Sec.~\ref{sec:AVCP}
by formulating the~AVCP. In Sec.~\ref{sec:deduction}, the AVCP is
used to obtain several generalized operator rules which connect the
average values of operators at different times, from which the
commonly used operator rules of quantum theory follow as a special
case. Using the AVCP and Postulate~3.4~(from Paper~I), we then
derive the explicit form of the temporal evolution operator, which
completes the derivation of the finite-dimensional abstract quantum
formalism begun in Paper~I.

Next, in Sec.~\ref{sec:commutation-relations}, taking the
infinite-dimensional form of the abstract quantum formalism as a
given, we use the AVCP to derive many of the commonly employed
formal rules of quantum theory, namely~(a) the commutation
relations~$[\cmatrix{x},\cmatrix{p}_x]=i\hbar$ and~$[\cmatrix{L}_x,
\cmatrix{L}_y]=i\hbar\cmatrix{L}_z$, and Dirac's Poisson bracket
rule, (b)~the explicit form of the operators for displacements and
rotations, and (c)~the relation between momentum and displacement
operators, and between angular momentum and rotation operators.

Finally, in Sec.~\ref{sec:arbitrariness}, we show that, in the
functions~$f(\chi_i) = \pm\cos(a\chi_i + b)$
and~$\tilde{f}(\chi_i) = \pm\sin(a\chi_i + b)$ derived in Paper~I,
the signs can be taken to be positive and the constants~$a$
and~$b$ can be taken to have the values~$a=1$ and~$b=0$ without
loss of generality.

We note that the treatment of the formal rules is illustrative
rather than exhaustive, so that only the most commonly encountered
measurement and transformation operators which are needed to
formulate non-relativistic and relativistic quantum mechanics have
been discussed. Many other formal rules~(such as the operators for
Galilei transformations, temporal displacement, and discrete
transformations such as spatial inversion) can be obtained by
arguments which closely follow those presented. The paper concludes
with a discussion of the results obtained.

\section{The Average-Value Correspondence Principle}
\label{sec:AVCP}

\subsection{Introduction}
\label{sec:AVCP-introduction}

Suppose that, as described in Paper~I, a quantum
model~$\model{q}(N)$, of dimension~$N$, has been constructed to
describe an abstract experimental set-up consisting of a source of
identical systems, a measurement set~$\set{A}$, and an interaction
set~$\set{I}$.  The measurements in~$\set{A}$, and the degenerate
forms of measurements in~$\set{A}$~\footnote{
A measurement that is a degenerate form of a
measurement~$\mment{A}$ is defined operationally in Paper~I.  Such
a measurement has~$N'<N$ possible outcomes, and can be represented
as an $N$-dimensional degenerate Hermitian operator~(with~$N'$
distinct eigenvalues).},
are represented by $N$-dimensional Hermitian operators~(possibly
with degenerate eigenvalues), and shall be called \emph{quantum}
measurements.

Suppose that quantum measurement~$\mment{A}$, with
operator~$\cmatrix{A}$, represents a measurement that is classically
described as a measurement of some observable~$A$~(which we shall
henceforth abbreviate to~``quantum measurement~$\mment{A}$~(or
operator~$\cmatrix{A}$) represents a measurement of~$A$''). Suppose
that we wish to determine whether there is a quantum measurement
that represents a measurement of, say,~$A^2$ and, if so, to
determine the operator which can be said to represent this
measurement. One can imagine that, classically, a measurement
of~$A^2$ is implemented by a process where a measurement of~$A$ is
performed and the outcome is then squared. If this implementation is
described in the quantum model, it follows that, if the input state
is an element,~$\cvect{v}_i$, of an orthonormal set of eigenvectors
of~$\cmatrix{A}$, the output state of the process is~$\cvect{v}_i$
and the observed result of the process is~$a_i^2$,
where~$\cmatrix{A}\cvect{v}_i = a_i \cvect{v}_i$~($i=1,2, \dots,
N$).  According to this line of reasoning, it follows at once that
the operator~$\cmatrix{A}^2$ therefore represents a measurement
of~$A^2$.

However, this simple argument does not readily generalize.  For
example, suppose that quantum measurements~$\mment{A}$
and~$\mment{B}$, with operators~$\cmatrix{A}$ and~$\cmatrix{B}$,
represent measurements of~$A$ and~$B$, respectively.  Suppose that
we wish to determine the operator~(if any exists) that represents a
measurement of~$A+B$. Classically, one imagines that such a
measurement is implemented by making measurements of~$A$ and~$B$
simultaneously on a system, and adding the outcomes. However,
if~$[\cmatrix{A}, \cmatrix{B}] \ne 0$, this implementation cannot be
described without modification in the quantum framework since the
measurements cannot be performed at the same time, and the order in
which they are performed is of potential significance.

Now, the conventional operator rules of quantum theory assert that a
measurement of~$A+B$ is represented by the
operator~$\cmatrix{A}+\cmatrix{B}$. Although this seems entirely
reasonable on a formal, symbolic level, it is not clear in what
physical sense~$\cmatrix{A}+\cmatrix{B}$ can be said to `represent'
the measurement since, whereas the eigenvectors and eigenvalues
of~$\cmatrix{A}$ directly reflect the output states and outcome
values obtained when measurement~$\mment{A}$ is performed on the
system, the eigenvectors of~$\cmatrix{A}+\cmatrix{B}$ do not, in
general, coincide with the eigenvectors of either~$\cmatrix{A}$
or~$\cmatrix{B}$, and the eigenvalues of~$\cmatrix{A}+\cmatrix{B}$
do not, in general, coincide with the possible results of any
plausible implementation~(described in the quantum model) of a
measurement of~$A+B$ that involves performing measurements of~$A$
and~$B$ and combining their outcomes. For example, a measurement
represented by operator~$\cmatrix{S}_x + \cmatrix{S}_z$ on a
spin-1/2 particle has possible outcomes~$\pm \hbar/\sqrt{2}$,
whereas the possible results of an implementation where a
measurement of~$S_x$ is followed by a measurement of~$S_z$~(or
vice-versa), and their outcomes are added, are~$\pm \hbar, 0$.

The above observations illustrate the difficulty of obtaining a
physical understanding of the operator rules of quantum theory even
in simple cases of interest.   Below, we shall formulate a physical
principle which gives a clear physical meaning to the sense in which
an operator can be said to represent a classically-described
measurement, and, in the majority of cases of interest, uniquely
determines the operator which represents such a measurement.

\subsubsection{Implementations of classically-described
measurements}

\begin{figure}[!h]
\begin{centering}
\includegraphics[width=3.25in]{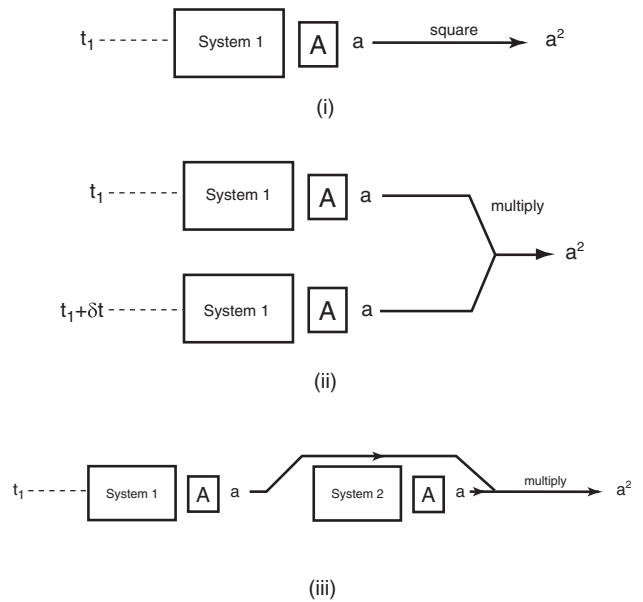}
\caption{\label{fig:mments-of-A-squared} Three implementations of
the measurement classically described as ``a measurement of~$A^2$''.
In~(i), a measurement of~$A$ is made on one copy of the system, and
the outcome is squared to give the result; in (ii), two immediately
successive measurements of~$A$ are made on one copy of the system,
and the result is obtained by multiplying the two outcomes; and
(iii)~two simultaneous measurements of~$A$ are made on two separate
copies of the system prepared in the same state, and the result is
obtained by multiplying the two outcomes. In a classical model of
this situation, each implementation yields the same result. However,
in the quantum model of these implementations, the expected result
of~(i) and~(ii) is~$\overline{a^2}$ whereas the expected result
of~(iii) is~$(\overline{a})^2$.}
\end{centering}
\end{figure}

Consider again a measurement that, from the classical standpoint, is
said to be a measurement of~$A^2$, where~$A$ is some observable.
Classically, one can imagine a measurement of~$A^2$ being
implemented in one of three different ways~(see
Fig.~\ref{fig:mments-of-A-squared}): (i)~make a measurement of~$A$
on one copy of the system, and square the outcome; (ii)~make two
immediately successive measurements of~$A$ on one copy of the
system, and multiply the two outcomes; or~(iii)~make two
simultaneous measurements of~$A$ on two separate copies of the
system prepared in the same state, and multiply the two outcomes.
Although the first of these implementations is the one we have
considered above, all three implementations yield the same result
when modeled classically, and so can be regarded as equally valid
implementations of a measurement of~$A^2$.

Now, perhaps surprisingly, when described using the quantum model,
these implementations do not, in general, yield the same expected
results. Consider a quantum experimental arrangement employing the
first implementation. In each run of the experiment, one copy of the
system is prepared in some given state. Let the probability that a
measurement of~$A$ yields outcome~$i$~($i=1, 2, \dots, N$), with
value~$a_i$, be denoted~$P_i$.  Then, the expected result is given
by
\begin{equation}
\begin{split}
\label{eqn:expected-outcome-example-1-implementation-1}
\text{Expected result} &= \sum_i (a_i)^2 P_i \\
                                &= \overline{a^2}
\end{split}
\end{equation}
One finds that implementation~(ii) yields the same expected
result. However, in implementation~(iii), where, in each run of
the experiment, two copies of the system are prepared in the same
state, one obtains
\begin{equation}
\label{eqn:expected-outcome-example-1-implementation-3}
\begin{split}
\text{Expected result} &= \sum_{i,j} (a_i a_j) P_iP'_j \\
                                &= (\overline{a})^2,
\end{split}
\end{equation}
with~$P_i$ and~$P'_j$ denoting respectively the probabilities that
the measurements of~$A$ on the first and second copy yield
outcomes~$i$ and~$j$~($i, j=1, 2, \dots, N$).

In this example, the differences between the implementations as
viewed in the quantum model arise due to the fact that, in the
quantum framework, an immediate repetition of a measurement on the
same copy of a system is different from performing a second
simultaneous measurement on a separate, identically-prepared, copy
of the system.

\begin{figure}[!h]
\begin{centering}
\includegraphics[width=3.25in]{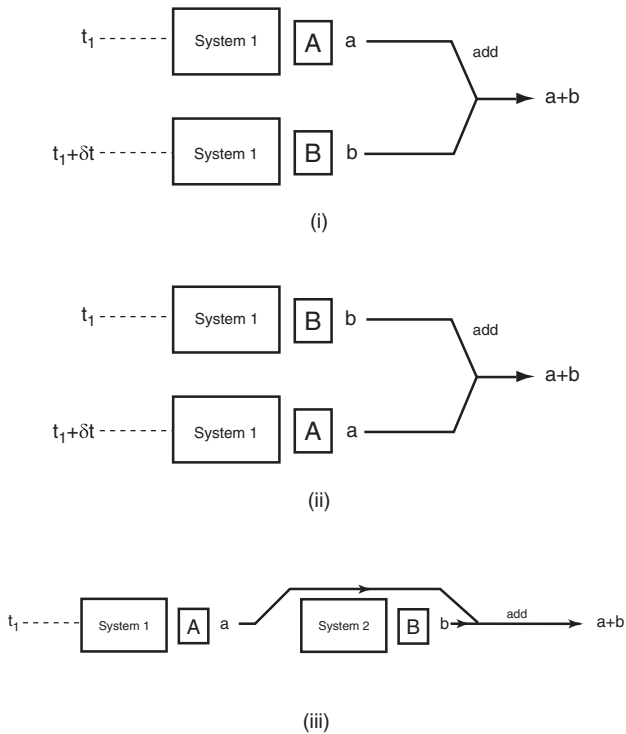}
\caption{\label{fig:mments-of-A-plus-B} Three implementations of the
measurement classically described as ``a measurement of~$A+B$''.
In~(i), a measurement of~$A$, and then a measurement of~$B$, is made
on one copy of the system, and the outcomes are then added to give
the result; in (ii), a measurement of~$B$, and then a measurement
of~$A$, is made on one copy of the system, and the outcomes are then
added to give the result; and (iii)~simultaneous measurements of~$A$
and~$B$ are made on two separate copies of the system prepared in
the same state, and the result is obtained by adding the two
outcomes. In a classical model of this situation, each
implementation yields the same result. However, in the quantum model
of this situation, the three implementations do not, in general,
yield the same expected results.}
\end{centering}
\end{figure}

In the case of a measurement that is classically described as a
measurement of~$A+B$, where measurements of~$A$ and~$B$ are assumed
to occur at the same time, one needs to take into account the
additional fact that, in the quantum framework, the order in which
two measurements is performed is of possible importance.
Accordingly, one can imagine implementing a measurement of~$A+B$ in
one of at least three different ways~(see
Fig.~\ref{fig:mments-of-A-plus-B}): (i)~make a measurement of~$A$,
and then a measurement of~$B$, on one copy of the system, and add
the outcomes; (ii) make a measurement of~$B$, and then a measurement
of~$A$, on one copy of the system, and add the outcomes; (iii)~make
a measurement of~$A$ on one copy of the system and, simultaneously,
a measurement of~$B$ on a second copy of the system prepared in the
same state as the first copy, and add the outcomes. Once again, in a
classical model of this situation, the results agree. However, if
one calculates the expected results in the quantum model, one finds
that, if~$[\cmatrix{A}, \cmatrix{B}]\ne 0$, all three will, in
general, disagree with one another.

As these examples illustrate, different implementations of the same
classically-described measurement do not, in general, give the same
expected result in the quantum model of the experimental
arrangement. Thus, whereas a classical description of a measurement,
such as ``a measurement of~$x^2$'', adequately describes the
intended measurement insofar as the outcome is concerned in the
classical framework, such a description allows for more than one
implementation which, in general, do not yield the same expected
result when modeled in the quantum framework.

\subsubsection{The average-value condition}
\label{sec:av-condition}

The existence of different implementations of the same
classically-described measurement immediately raises two questions.
First, are these implementations, in some sense, equally valid in
the quantum framework, or it is possible to find some reasonable
physical basis upon which to select particular implementations and
regard these as more fundamental than the others? Second, is it
possible to find operators that represent the selected
implementations, and, if so, do all the selected implementations of
a given measurement have the same operator representation?

To answer these questions, we begin by observing that, although the
above implementations are all regarded as \emph{bone fide}
measurements in the classical model, a measurement is only
describable as such in the quantum model, and so can be called a
quantum measurement, if it can be represented by a Hermitian
operator which represents a single measurement performed upon one
copy of the system at a particular time. So, for example, although
implementation~(iii) of a measurement of~$A^2$ can be modeled in the
quantum framework, the process \emph{as a whole} cannot be described
as a quantum measurement since it involves two separate
measurements. In contrast, implementation~(i) of a measurement
of~$A^2$ can be described as a quantum measurement since it only
involves a single measurement on one copy of the system.

However, although an implementation modeled in the quantum framework
that involves more than one measurement cannot itself be regarded as
a quantum measurement, we can reasonably ask whether it is possible
to find a quantum measurement,~$\mment{C}$, with
operator~$\cmatrix{C}$, which, in some sense to be determined, can
nonetheless be said to \emph{represent} the implementation.

At the outset, we note that it makes no sense to require that
measurement~$\mment{C}$ yield the same outcome as a given
implementation since measurement outcomes are only probabilistically
determined in the quantum framework. However, we \emph{can} impose
the simple condition that, over an infinite number of runs,  the
average value of measurement~$\mment{C}$ should coincide with the
average result obtained from the implementation for any initial
state of the system.

For example, in the case of an implementation of a measurement
of~$A^2$, measurement~\mment{C} which, by hypothesis, represents the
implementation, must be such that, for all states of the
system,~$\ev{\cmatrix{C}}$ is equal to the expected result obtained
from the implementation.  In the case of implementations~(i)
and~(ii) described above, using
Eq.~\eqref{eqn:expected-outcome-example-1-implementation-1}, we
accordingly obtain the condition that the relation
\begin{equation}
\begin{split}
\ev{\cmatrix{C}}    &= \overline{a^2} \\
                    &= \ev{\cmatrix{A}^2}
\end{split}
\end{equation}
must hold for all states,~$\cvect{v}$, of the system.  From this
condition, we can conclude that
\begin{equation}
\label{eqn:avc-rule-1}
\cmatrix{C} = \cmatrix{A}^2.
\end{equation}
In the case of
implementation~(iii), using
Eq.~\eqref{eqn:expected-outcome-example-1-implementation-3}, we
obtain the condition that the relation
\begin{equation}
\begin{split}
\ev{\cmatrix{C}}    &= (\overline{a})^2 \\
                    &= \ev{\cmatrix{A}}^2
\end{split}
\end{equation}
must hold for all~$\cvect{v}$.  By diagonalizing~$\cmatrix{A}$, one
can readily show that this condition implies that~$\cmatrix{A}$ is a
multiple of the identity, which represents a trivial measurement
that yields the same outcome irrespective of the state of the
system. Therefore, implementation~(iii) does not satisfy the above
average-value condition in the case of any non-trivial measurement
of~$A$, and can therefore be reasonably eliminated. Hence, in this
case, the average-value condition is sufficiently strong so as to be
able to pick out implementations~(i) and~(ii), and, since the
average-value condition also implies that these implementations are
both represented by the operator,~$\cmatrix{A}^2$, one can
unambiguously conclude that a measurement of~$A^2$ is represented by
the operator~$\cmatrix{A}^2$.

Proceeding in a similar way, restricting ourselves for the time
being to measurements~$\mment{A}$ and~$\mment{B}$ that are not
sub-system measurements~\footnote{A sub-system measurement is
defined in Paper~I as a measurement performed on a single sub-system
of a composite system.}, one finds that, in the case of a
measurement of~$C = A+B$, only implementation~(iii) is possible
if~$[\cmatrix{A}, \cmatrix{B}]\ne 0$, which then yields the operator
\begin{equation}
\label{eqn:avc-rule-2} \cmatrix{C} = \cmatrix{A}+\cmatrix{B}.
\end{equation}
If~$[\cmatrix{A}, \cmatrix{B}]= 0$, then all three implementations
are possible, and all yield the same operator,~$\cmatrix{C}$, as
above.

Finally, in the case of a measurement of~$AB$, with~$[\cmatrix{A},
\cmatrix{B}]=0$, one finds that the implementation must the one
where measurements~\mment{A} and~\mment{B} are performed on the same
copy of the system, in which case the
operator~$\cmatrix{A}\cmatrix{B}$ is obtained.

Hence, we see that the average-value condition is sufficient to
yield a unique operator representation for the measurements
considered above.  Based on the above considerations, we can
tentatively formulate the following general rule:~in the case of a
measurement which has an implementation that contains two elementary
measurements~(not sub-system measurements) represented by commuting
operators, it is possible to find a quantum measurement that
represents the implementation if the two elementary measurements are
performed on the same copy of the system; but, when the operators do
not commute, the elementary measurements must be performed on
different copies of the system.

We now consider implementations of a measurement of~$AB$ in the case
when~$[\cmatrix{A}, \cmatrix{B}]\ne 0$. The general rule just given
suggests that we should consider the implementation of this
measurement where the measurements of~$A$ and~$B$ are performed on
different copies of the system. This implementation has the expected
value
\begin{equation}
\sum_i \sum_j (a_i b_j) P_i P_j' =
\ev{\cmatrix{A}}\ev{\cmatrix{B}},
\end{equation}
with~$P_i$ and~$P'_j$ denoting respectively the probabilities that
the measurement of~$A$ on the first copy and the measurement
of~$B$ on the second copy yield outcomes~$i$ and~$j$~($i, j=1, 2,
\dots, N$), and~$a_i, b_j$ respectively denoting the values of the
$i$th and $j$th outcomes of measurements~$\mment{A}$
and~$\mment{B}$. Imposing the above average-value condition, the
operator~$\cmatrix{C}$ that represents this implementation must
satisfy the condition
\begin{equation} \label{eqn:measurement-AB-av-desideratum}
\ev{\cmatrix{C}} =  \ev{\cmatrix{A}}\ev{\cmatrix{B}}
\end{equation}
for all~$\cvect{v}$.  However, for non-commuting~$\cmatrix{A}$
and~$\cmatrix{B}$, one finds that~$\cmatrix{C}$ cannot be found
such that this relation is satisfied for all~$\cvect{v}$. We note,
however, that with
\begin{equation}
\label{eqn:hermitized-form-AB} \cmatrix{C} = \frac{1}{2}
(\cmatrix{A}\cmatrix{B} + \cmatrix{B}\cmatrix{A}),
\end{equation}
equation~\eqref{eqn:measurement-AB-av-desideratum} holds for the
eigenstates of~$\cmatrix{A}$ and the eigenstates of~$\cmatrix{B}$,
which suggests that we may be able to weaken the average-value
condition in this case so that we only require that
Eq.~\eqref{eqn:measurement-AB-av-desideratum} holds for these
eigenstates.  However, as we shall illustrate
later~(Sec.~\ref{sec:inconsistencies}), the application of
Eq.~\eqref{eqn:hermitized-form-AB} can lead to inconsistencies.
Consequently, we conclude that, from the point of view of the
average-value condition, it is not possible to find an operator that
represents a measurement of~$AB$ when the operators~$\cmatrix{A}$
and~$\cmatrix{B}$ do not commute. More generally, we find that,
when~$[\cmatrix{A}, \cmatrix{B}]\ne 0$, it is necessary to exclude
measurements of~$f(A,B)$, with~$f$ analytic~(so that~$f$ has a
well-defined polynomial expansion in~$A$ and~$B$), where the
polynomial expansion of~$f$ contains product terms.

Finally, in the case of a measurement which has a classical
implementation which contains two elementary measurements that are
performed on different sub-systems of a composite system, one finds
that the implementation can be represented by a quantum measurement
irrespective of whether the two measurements are performed on the
same or on different copies of the system, and that the different
possible implementations are represented by the same operator.

\subsubsection{Generalizations}

In the examples above, we have considered implementations of a
classically-described measurement in which the measurement and the
elementary measurements in the implementation are performed at the
same time, and in the same frame of reference. However, as the
following examples illustrate, these are unnecessary restrictions.

First, classically, for a non-relativistic particle of mass~$m$, one
can implement a measurement of~$x$ performed at time~$t+\delta t$ by
a process where measurements of~$x$ and~$p_x$ are performed at
time~$t$, and the function~$x + p_x \,\delta t/m$ is then computed.

Second, one can implement a measurement of~$x'$ on a particle in the
reference frame~$S'$ that is displaced along the $x$-axis relative
to frame~$S$ by performing a measurement of~$x$ in frame~$S$, and
computing the appropriate function~$x'=f(x)$ that relates~$x$
and~$x'$.

The above considerations regarding the implementation of a
classically-described measurement are applicable without change to
the case where the measurement and the elementary measurements in
its implementation are performed at different times or in
different frames of reference. Below, we shall accordingly
generalize our notion of the implementation of a
classically-described measurement.

\subsection{Statement of the Principle}
\label{eqn:AVCP-statement}

\begin{figure}[!h]
\begin{centering}
\includegraphics[width=3.25in]{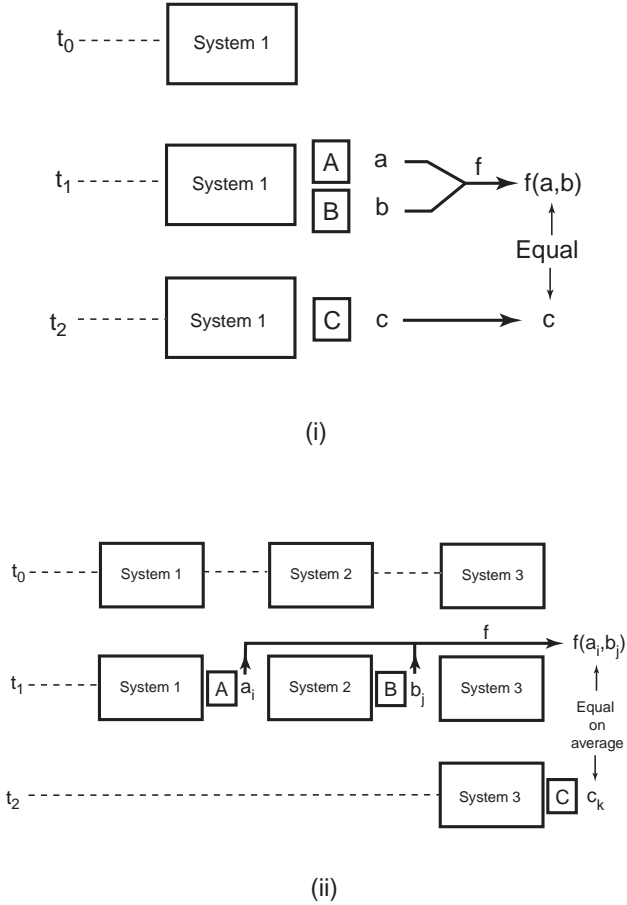}
\caption{\label{fig:AVCP-example} An example of the application of
the AVCP.  (i) A classical experiment showing the measurements
of~$A, B$ and~$C$ performed at times~$t_A, t_B$, and~$t_C$,
respectively, with outcomes denoted as~$a, b$, and~$c$,
respectively. Here,~$t_A = t_B = t_1$ and~$t_C = t_2$. Suppose that
one finds that the relation~$c=f(a,b)$ holds for all initial states
of the system. (ii) The corresponding quantum experiment. Three
copies of the system are prepared in the same initial
state,~$\cvect{v}_0$, at time~$t_0$, and are placed in identical
backgrounds.  In this example, it is assumed that the
operators~$\cmatrix{A}$ and~$\cmatrix{B}$ do not commute. Hence, by
the AVCP, measurements~\mment{A} and~\mment{B} are performed on
different copies of the system. Measurement~\mment{C} is performed
on a separate copy of the system.  In any given run of the
experiment, the probabilities that measurements~\mment{A}, \mment{B}
and~\mment{C} yield outcome values~$a_i, b_j$ and~$c_k$~($i, j, k =
1, \dots, N$), are~$P_i, P'_j$ and~$P_k''$, respectively. The AVCP
then asserts that, provided the polynomial expansion of~$f(a,b)$
contains no product terms involving~$a$ and~$b$, the
relation~$\overline{c}=\overline{f(a,b)}$ holds for all initial
states,~$\cvect{v}_0$, of the system, where the average is taken
over an infinite number of runs of the experiment; that is, $\sum_k
c_k P_k'' = \sum_{ij} f(a_i,b_j) P_i P'_j$ for all~$\cvect{v}_0$.}
\end{centering}
\end{figure}

We shall now state a general principle which incorporates the
observations made above concerning the average-value condition. An
illustrative example is given in Fig.~\ref{fig:AVCP-example}.

\begin{itemize}
   \item[]\textbf{Average-Value Correspondence Principle}
    Consider a classical idealized experiment in which a system~(possibly
    a composite system) is
    prepared in some state at time~$t_0$, and is allowed to evolve
    in a given background. Suppose that a
    measurement of~$A^{(m)}$~$(m\ge 2)$, performed on the
    system at time~$t_2$ with outcome~$a^{(m)}$, has a classical
    implementation in which measurements of~$A^{(1)},
    A^{(2)}, \dots, A^{(m-1)}$ are performed upon one copy of the system
    at time~$t_1$, and their respective outcomes,
    denoted~$a^{(1)}, a^{(2)}, \dots,
    a^{(m-1)}$, are then used to compute the result~$f(a^{(1)},
    a^{(2)}, \dots, a^{(m-1)})$, where~$f$ is an analytic
    function, so that the relation
    \begin{equation*}
    \label{eqn:avc}
    a^{(m)}=f(a^{(1)}, a^{(2)}, \dots, a^{(m-1)}) \tag{$*$}
    \end{equation*}
    holds for all initial (classical)~states
    of the system.

    Consider the case where the quantum measurements~$\mment{A}^{(1)}, \mment{A}^{(2)}, \dots,
    \mment{A}^{(m)}$, with operators~$\cmatrix{A}^{(1)}, \cmatrix{A}^{(2)},
    \dots, \cmatrix{A}^{(m)}$, represent the
    measurements of~$A^{(1)}, A^{(2)}, \dots, A^{(m)}$, respectively.  Then, consider
    the following idealized quantum experimental
    arrangement consisting of several set-ups, each consisting of identical
    sources and backgrounds, where, in each set-up, a copy of the system is
    prepared in the same initial state,~$\cvect{v}_0$, at time~$t_0$.

    In one set-up, only measurement~$\mment{A}^{(m)}$ is
    performed~(at time~$t_2$) and, for any~$i,j$ with~$i\neq j$
    and~$i,j \leq m-1$, the measurements~$\mment{A}^{(i)},
    \mment{A}^{(j)}$ are performed~(at time~$t_1$) in the same
    set-up in the quantum experiment if~$\bigl[\cmatrix{A}^{(i)},
    \cmatrix{A}^{(j)}\bigr]=0$ provided that, if the system is
    composite, the measurements are performed on the same
    sub-system; and are performed~(at time~$t_1$) in different
    set-ups if~$\bigl[\cmatrix{A}^{(i)},
    \cmatrix{A}^{(j)}\bigr]\ne 0$.

    Let the outcomes of the measurements~$\mment{A}^{(1)}, \dots,
    \mment{A}^{(m)}$ in any given run of the experimental arrangement be
    denoted~$a^{(1)}, \dots, a^{(m)}$, respectively.
    The function~$f(a^{(1)}, a^{(2)}, \dots, a^{(m-1)})$
     is defined
    as \emph{simple} provided that its
    polynomial expansion contains no terms involving a product of
    eigenvalues belonging to measurements whose operators
    do not commute.  If~$f$ is simple, then~\eqref{eqn:avc} holds
    on average, the average being taken over an infinitely large
    number of runs of the experiment.

\end{itemize}

The above principle can be generalized in a number of ways, for
example to the case where the measurements of~$A^{(1)}, \dots,
A^{(m-1)}$ are not performed at the same time. However, these
generalizations are unnecessary for the derivations of the usual
formal rules of quantum theory, and are therefore not discussed
here.

\section{Generalized Operator Rules and the Temporal Evolution Operator}
\label{sec:deduction}

\subsection{Generalized Operator Rules}

We will now apply the AVCP to derive operator relations which hold
when the function~$f$ takes various useful forms.  In each
instance of~$f$, we shall first derive a generalized operator rule
which relates the expected values of the relevant operators at
\emph{different times}.  Then, taking the special case when the
expected values are computed at the same time, we obtain the
corresponding operator rule which relates the operators
themselves.

We shall consider a classical experiment where a system is subject
to measurements of~$A$ and~$B$ at time~$t_1$, and to a measurement
of~$C$ at time~$t_2$. We shall suppose that measurement of~$C$, with
outcome~$c$, can be implemented by an arrangement in which the
measurements of~$A$ and~$B$ are performed, with respective
outcomes~$a$ and~$b$, and the function~$f(a,b)$ then computed, so
that the relation
\begin{equation}
c= f(a,b)
\end{equation}
holds for all initial states of the system.

In a quantum model of the appropriate experimental arrangement, let
the operators that represent these measurements be
denoted~$\cmatrix{A}$,~$\cmatrix{B}$, and~$\cmatrix{C}$,
respectively.  To simplify the presentation, we shall only consider
the case where these operators have finite dimension,~$N$; the
results obtained below can be readily shown to apply in the infinite
dimensional case. Let the elements of orthonormal sets of
eigenvectors of~$\cmatrix{A}, \cmatrix{B}$ and~$\cmatrix{C}$ be
denoted~$\cvect{v}_i, \cvect{v}_j'$, and~$\cvect{v}''_k$,
respectively~$(i, j, k = 1,2, \dots, N)$, let the corresponding
eigenvalues be denoted~$a_i, b_j$ and~$c_k$, and let the
probabilities of the~$i$th,~$j$th and~$k$th outcomes of
measurements~$\mment{A}$,~$\mment{B}$ and~$\mment{C}$ in any given
experimental arrangement be denoted by~$P_i$,~$P'_j$ and~$P_k''$,
respectively.

\subsubsection*{Case 1. $f$ is a function of~$a$ only.}
\label{sec:AVCP-case1}

In this case, the quantum experiment simply consists of two
identical set-ups, involving two copies of the system,
where~$\mment{A}$ is performed on one copy at time~$t_1$
and~$\mment{C}$ on the other copy at time~$t_2$. Since
function~$f$ is simple, by the AVCP, the relation
\begin{equation} \label{eqn:AVCP-case-1-main-eqn}
    \sum_k c_k P_k'' =  \sum_i f(a_i) P_i
\end{equation}
holds for all initial states,~$\cvect{v}_0$, of the system.
Explicitly,
\begin{equation}
\begin{aligned}
P_i &= \big|\cvect{v}_i^\dagger \cvect{v}_{t_1}\big|^2 \\
P_k'' &= \big|\cvect{v}_k''^\dagger \cvect{v}_{t_2}\big|^2,
\end{aligned}
\end{equation}
with~$\cvect{v}_t$ being the state of the relevant copy of the
system at time~$t$.  Hence, we can write
Eq.~\eqref{eqn:AVCP-case-1-main-eqn} as
\begin{equation}
\cvect{v}^\dagger_{t_2}
                        \bigg(
                        \sum_k \cvect{v}_k'' {\cvect{v}''_k}^\dagger
                        c_k
                        \bigg)
\cvect{v}_{t_2} = \cvect{v}^\dagger_{t_1} \bigg(
                        \sum_i \cvect{v}_i \cvect{v}_i^\dagger f(a_i)
                        \bigg)
\cvect{v}_{t_1}.
\end{equation}
Noting that
\begin{equation}
\begin{aligned}
f(\cmatrix{A}) &= \sum_i \cvect{v}_i \cvect{v}_i^\dagger f(a_i) \\
\cmatrix{C} &= \sum_k \cvect{v}_k'' {\cvect{v}''_k}^\dagger c_k,
\end{aligned}
\end{equation}
we obtain the relation
\begin{equation} \label{eqn:AVCP-case-1-main-result}
\ev{\cmatrix{C}}_{t_2} = \evb{f(\cmatrix{A})}_{t_1},
\end{equation}
which holds for all~$\cvect{v}_0$.  We can summarize the above
result in the form of the \emph{generalized function rule}:
\begin{equation} \label{eqn:AVCP-case-1-general-rule}
c(t_2) = f\left(a(t_1)\right) \mapsto \ev{\cmatrix{C}}_{t_2} =
\evb{f(\cmatrix{A})}_{t_1} \quad\forall \cvect{v}_0,
\end{equation}
where, for clarity, the times at which the outcomes are obtained has
been explicitly indicated.
In the special case where~$t=t_1 = t_2$, we obtain the usual
operator rule, the \emph{function rule}:
\begin{equation} \label{eqn:AVCP-case-1-special-case-result}
c = f(a) \mapsto  \cvect{C} =  f(\cvect{A}).
\end{equation}

\subsubsection*{Case 2. $f(a,b)= f_1(a) +  f_2(b)$}
\label{sec:AVCP-case2}

It is necessary to consider three sub-cases.  First, if
measurements~\mment{A} and~\mment{B} have commuting operators and,
in the case of a composite system, if they are sub-system
measurements performed on the same sub-system, then, by the AVCP,
they are performed on the same copy of the system in the quantum
experiment. Since~$f$ is simple, the AVCP applies, so that
\begin{equation}
\begin{split}
    \sum_k c_k P_k''    &=  \sum_i \bigg( f_1(a_i)  + \sum_j  f_2(b_j)
                                                             P'_{j|i}
                                    \bigg) P_i \\
                        &=   \sum_i \left( f_1(a_i)  +  f_2(b_i)
                                    \right) P_i,
\end{split}
\end{equation}
holds for all initial states,~$\cvect{v}_0$, of the system.  Here,
the notation~$P'_{j|i}$ is the probability that
measurement~\mment{B} yields outcome~$j$ given that~\mment{A} has
yielded outcome~$i$; in this case,~$P'_{j|i} = \delta_{ij}$.  From
the above relation, we obtain the generalized operator relation
\begin{equation} \label{eqn:AVCP-case2-av-relation}
\ev{\cmatrix{C}}_{t_2} = \evb{f_1(\cmatrix{A})}_{t_1} +
\evb{f_2(\cmatrix{B})}_{t_1},
\end{equation}
which holds for all initial states,~$\cvect{v}_0$.  In the special
case where~$t_1=t_2$, we obtain the operator relation,
\begin{equation} \label{eqn:C-operator1}
    \cmatrix{C} =  f_1(\cmatrix{A}) +  f_2(\cmatrix{B}).
\end{equation}

Second, in the case where measurements~\mment{A} and~\mment{B} are
sub-system measurements performed on different sub-systems, they
can, by the AVCP, be performed on the same copy of the system, in
which case we obtain the same results as above.

Third, if measurements~\mment{A} and~\mment{B} have non-commuting
operators, then, by the AVCP, they are performed on different
copies of the system in the quantum  experiment. Since~$f$ is
simple, the AVCP again applies, so that the relation
\begin{equation}
    \sum_k c_k P_k''    = \sum_i   f_1(a_i) P_i + \sum_j f_2(b_j) P_j'
\end{equation}
holds for all initial states of the system, which yields the same
relation as in Eq.~\eqref{eqn:AVCP-case2-av-relation}.

Hence, combining the foregoing three sub-cases, we obtain the
\emph{generalized sum rule}:
\begin{multline} \label{eqn:AVCP-case-2-general-rule}
c(t_2) = f_1\left(a(t_1)\right) + f_2\left(b(t_1)\right) \\
\mapsto \ev{\cmatrix{C}}_{t_2} = \evb{f_1(\cmatrix{A})}_{t_1} +
                        \evb{f_2(\cmatrix{B})}_{t_1} \quad\forall \cvect{v}_0.
\end{multline}
In the special case where~$t=t_1 = t_2$, we obtain the \emph{sum
rule}:
\begin{equation} \label{eqn:AVCP-case-2-special-case-result}
c = f_1(a) + f_2(b) \mapsto  \cvect{C} =  f_1(\cvect{A}) +
                                                f_2(\cvect{B}).
\end{equation}
More generally, consider a classical experiment where measurements
of~$A^{(1)}, A^{(2)}, \dots, A^{(m-1)}$ are performed on a system at
time~$t_1$ and a measurement of~$A^{(m)}$ at time~$t_2$, with
outcomes~$a^{(1)}, \dots, a^{(m)}$, respectively. Suppose that the
relation
\begin{equation} a^{(m)} = f(a^{(1)}, a^{(2)}, \dots, a^{(m-1)}),
\end{equation}
where%
\begin{equation} f(a^{(1)}, a^{(2)}, \dots, a^{(m-1)}) =
f_1(a^{(1)}) + \dots + f_{m-1}(a^{(m-1)}),
\end{equation}
holds in the classical model for all initial states, and that, in
the quantum model, the measurements are represented by the
operators~$\cmatrix{A}^{(1)}, \cmatrix{A}^{(2)}, \dots,
\cmatrix{A}^{(m)}$, respectively.  Then, one finds that the AVCP
implies the generalized operator rule:
\begin{multline} \label{eqn:AVCP-case2-av-relation-general-av}
a^{(m)}(t_2) = \sum_{l=1}^{m-1} f_l\bigl(a^{(l)}(t_1)\bigr)
\\
\mapsto \evb{\cmatrix{A}^{(m)}}_{t_2} = \sum_{l=1}^{m-1}
\evb{f_l(\cmatrix{A}^{(l)})}_{t_1} \quad \forall\cvect{v}_0.
\end{multline}
Taking the special case of simultaneous measurements~($t_1 =
t_2$), we obtain the operator rule
\begin{equation} \label{eqn:AVCP-case2-av-relation-general}
a^{(m)} = \sum_{l=1}^{m-1} f_l(a^{(l)}) \mapsto \cmatrix{A}^{(m)} =
\sum_{l=1}^{m-1} f_l(\cmatrix{A}^{(l)}).
\end{equation}

\subsubsection*{Case 3. $f(a,b)= f_1(a)f_2(b)$}
\label{sec:AVCP-case3}

We again consider three sub-cases. First, if measurements~\mment{A}
and~\mment{B} are represented by commuting operators and, if the
measurements are sub-system measurements and are performed on the
same sub-system, then, in the quantum experiment, they are performed
on the same copy of the system. Since~$f$ is, therefore, simple, the
AVCP applies, so that the relation
\begin{equation}
\begin{split}
    \sum_k c_k P_k''    &=  \sum_{i,j}  f_1(a_i)f_2(b_j)  P_i
                                                            P'_{j|i} \\
                        &=   \sum_i f_1(a_i) f_2(b_i) P_i
\end{split}
\end{equation}
holds for all initial states,~$\cvect{v}_0$, of the system.  Hence,
the generalized operator relation%
\begin{equation}
\ev{\cmatrix{C}}_{t_2}  =
\evb{f_1(\cmatrix{A})f_2(\cmatrix{B})}_{t_1}
\end{equation}
holds for all~$\cvect{v}_0$.  In the case where
measurements~\mment{A}, \mment{B}, and~\mment{C} are simultaneous,
we obtain the operator relation
\begin{equation}
    \cmatrix{C} = f_1(\cmatrix{A}) f_2(\cmatrix{B}).
\end{equation}

Second, if measurements~\mment{A} and~\mment{B} are represented by
commuting operators and are sub-system measurements performed on
different sub-systems of a composite system, then they can be
performed on the same copy of the system in the quantum
experiment, in which case we obtain the same result as above.

Third, if measurements~\mment{A} and~\mment{B} are represented by
non-commuting operators, then the function~$f$ is not simple, and
the AVCP does not apply.

We can combine the three foregoing sub-cases to obtain the
 \emph{generalized product
rule}:
\begin{multline} \label{eqn:AVCP-case-3-general-rule}
c(t_2) = f_1\bigl(a(t_1)\bigr)f_2\bigl(b(t_1)\bigr) \\
\mapsto
\ev{\cmatrix{C}}_{t_2} = \evb{f_1(\cmatrix{A})
                                f_2(\cmatrix{B})}_{t_1} \quad\forall
                                \cvect{v}_0\quad \text{if}~[\cmatrix{A},
                                \cmatrix{B}]=0.
\end{multline}
In the special case where~$t_1 = t_2$, we obtain the \emph{product
rule}:
\begin{equation} \label{eqn:AVCP-case-3-special-case-result}
c = f_1(a)f_2(b) \mapsto  \cmatrix{C} =  f_1(\cmatrix{A})
                                                f_2(\cmatrix{B})
                                                \quad \text{if}~[\cmatrix{A},
                                                            \cmatrix{B}]=0.
\end{equation}

\subsubsection{Some comments on inconsistencies}
\label{sec:inconsistencies}

As mentioned in Sec.~\ref{sec:av-condition}, if the average-value
condition is weakened to allow  a measurement of~$AB$ to be
represented by an operator in the case where~$[\cmatrix{A},
\cmatrix{B}]\ne 0$, one is lead to a rule that is often stated,
namely
\begin{equation} \label{eqn:hermitization-rule}
f_1(a)f_2(b) \mapsto \frac{1}{2} \bigl(f_1(\cmatrix{A})
    f_2(\cmatrix{B}) + f_1(\cmatrix{A})f_2(\cmatrix{B}) \bigr)
\end{equation}
However, using this rule, one finds that inconsistencies quickly
arise.  For example, one can first apply this rule to find that
the operator representing a measurement of~$AB$ is
\begin{equation} \label{eqn:hermitized-form-AB2}
\widehat{AB} = (\cmatrix{A}\cmatrix{B} +
\cmatrix{B}\cmatrix{A})/2,
\end{equation}
where the notation~$\widehat{X}$ is used to denote the operator that
represents a measurement of~$X$. One can then apply the rule a
second time to find the operator that represents a measurement
of~$A^2B$. By treating this measurement as a measurement of~$A(AB)$,
or as a measurement of~$(A^2)B$, one obtains, respectively, either
\begin{equation}
\begin{split}
\widehat{A(AB)} &=\frac{1}{2} (\cmatrix{A}\widehat{AB} +
                                \widehat{AB}\cmatrix{A}) \\
                &=\frac{1}{4}
                \left(\cmatrix{A}(\cmatrix{A}\cmatrix{B} +
                \cmatrix{B}\cmatrix{A}) +
                (\cmatrix{A}\cmatrix{B} +
                \cmatrix{B}\cmatrix{A})\cmatrix{A} \right) \\
                &=\frac{1}{4} \left(\cmatrix{A}^2\cmatrix{B} +
                2\cmatrix{A}\cmatrix{B}\cmatrix{A} +
                \cmatrix{B}\cmatrix{A}^2 \right),
\end{split}
\end{equation}
or
\begin{equation}
\begin{split}
\widehat{(A^2)B} &=\frac{1}{2} (\widehat{A}^2\cmatrix{B} +
                                \cmatrix{B}\widehat{A}^2) \\
                &= \frac{1}{2} (\cmatrix{A}^2\cmatrix{B} +
                                \cmatrix{B}\cmatrix{A}^2),
\end{split}
\end{equation}
which are, in general, inequivalent.  Hence, the average-value
condition cannot be applied to non-simple functions of
observables, even in weakened form, without leading to
inconsistencies.

We also remark that, given the AVCP, it cannot consistently be
maintained that every classically-described measurement is
represented by a quantum measurement, since, under this
assumption, the function and sum rules can be applied to a
measurement of~$(A+B)^2$, with~$[\cmatrix{A}, \cmatrix{B}]\ne 0$,
to derive Eq.~\eqref{eqn:hermitized-form-AB2} as follows. First,
defining~$d= a+b$, we use the sum rule to find~$\cmatrix{D} =
\cmatrix{A} + \cmatrix{B}$, and then use the function rule to find
that
\begin{equation}
\begin{split}
\widehat{D^2} &= \cmatrix{D}^2 \\
                &= \cmatrix{A}^2 + \cmatrix{A}\cmatrix{B} +
                    \cmatrix{B}\cmatrix{A} + \cmatrix{B}^2.
\end{split}
\end{equation}
Second, since, by assumption, a measurement of~$AB$ is represented
by a quantum measurement, we can use the sum rule directly to find a
measurement of~$D^2 = A^2 + 2AB + B^2$:
\begin{equation} \label{eqn:D-squared-2}
\begin{split}
\widehat{D^2} &= \widehat{A^2} + 2\widehat{AB} + \widehat{B^2} \\
                &= \cmatrix{A}^2 + 2\widehat{AB} + \cmatrix{B}^2.
\end{split}
\end{equation}
Equating these expressions for~$\widehat{D^2}$, we obtain
Eq.~\eqref{eqn:hermitized-form-AB2}, which, as we have seen, leads
to an inconsistency.   If the AVCP is accepted as valid, this
inconsistency can only be avoided if we conclude that a
measurement of~$AB$~(when~$[\cmatrix{A}, \cmatrix{B}]\ne 0$)
cannot be represented by a quantum measurement, in which case the
sum rule cannot be applied to obtain Eq.~\eqref{eqn:D-squared-2}.

In summary, given that measurements of~$A$ and~$B$ are represented
by quantum measurements~$\mment{A}$ and~$\mment{B}$, one can use the
AVCP to find quantum measurements that represent measurements
of~$f(A), f_1(A) + f_2(B)$ and, for~$[\cmatrix{A}, \cmatrix{B}]=0$,
of~$f_1(A)f_2(B)$; and, more generally, one can find quantum
measurements that represent measurements of~$f(A,B)$ when~$f$ is
simple. The AVCP also implies that a measurement of~$AB$
when~$[\cmatrix{A}, \cmatrix{B}]\ne 0$ cannot be represented by a
quantum measurement. However, this does not appear to be a
significant restriction since such measurements do not occur in the
Hamiltonian for a system of particles, in either non-relativistic or
relativistic physics, in which fundamental forces alone are acting.

\subsection{Temporal Evolution}

In this section, we will use the AVCP, together with
Postulate~3.4~(see Paper~I), to derive the explicit form of the
temporal evolution operator for a system in a time-dependent
background.

Temporal evolution of the system is represented by a unitary
transformation. Specifically, over the course of the interval~$[t,
t+\Delta t]$, the state~$\cvect{v}(t)$ evolves as
\begin{equation}
    \cvect{v}(t+\Delta t) = \cmatrix{U}_t(\Delta t) \cvect{v}(t),
\end{equation}
where~$\cmatrix{U}_t(\Delta t)$ is the unitary matrix that
represents temporal evolution of the system during~$[t, t+\Delta
t]$.

Suppose now that the background of the system is time-independent
during this interval. Then we shall write~$\cmatrix{U}_t(\Delta
t)$ as~$\cmatrix{V}_t(\Delta t)$. Now, for~$0 \leq \Delta t_1 +
\Delta t_2 \leq \Delta t$, and~$\Delta t_1, \Delta t_2$  both
positive, we have
\begin{equation}
    \cmatrix{V}_t(\Delta t_1 + \Delta t_2)
        = \cmatrix{V}_{t+\Delta t_1}(\Delta t_2)
          \cmatrix{V}_t(\Delta t_1).
\end{equation}
But the time-independence of the background implies
that~$\cmatrix{V}_{t+\Delta t_1}(\Delta t_2) =
\cmatrix{V}_{t}(\Delta t_2)$.  Therefore,
\begin{equation}
\cmatrix{V}_t(\Delta t_1 + \Delta t_2)
        = \cmatrix{V}_{t}(\Delta t_2)
          \cmatrix{V}_t(\Delta t_1),
\end{equation}
which can be solved to yield
\begin{equation}
\cmatrix{V}_t(\Delta t_1) = \exp(-i\cmatrix{K}_t \Delta t_1),
\end{equation}
where~$\cmatrix{K}_t$ is a Hermitian matrix.

To determine the nature of~$\cmatrix{K}_t$, we proceed as follows.
In the classical model of a physical system in a time-independent
interval~$[t, t+\Delta t]$, the classical Hamiltonian for the
system is not explicitly dependent upon time during this interval.
On the assumption that the classical Hamiltonian is a simple
function of the observables of the system, and that the
measurements of these observables are represented by quantum
measurements~
\footnote{If the measurements of the observables of which the
classical Hamiltonian is a function are not represented by quantum
measurements, then it is not possible to write down the operator
corresponding to the classical Hamiltonian.  Similarly, if the
classical Hamiltonian is \emph{not} a simple function of the
observables of the system, then the form of the operator that
represents a measurement of energy cannot be determined by the
AVCP and, therefore, the explicit form of~$\cmatrix{U}_t(\Delta
t)$ cannot be obtained from the argument given in the text.
However, these assumptions do not appear to represent a
significant restriction in any fundamental cases of interest~(in
both non-relativistic and relativistic cases).},
it follows from the AVCP that the corresponding Hamiltonian
operator is also not explicitly dependent upon time during this
interval, so that, in particular,
\begin{equation} \label{eqn:Hamiltonian-constraint-1}
\cmatrix{H}_t = \cmatrix{H}_{t+\Delta t},
\end{equation}
where~$\cmatrix{H}_t$ denotes the Hamiltonian operator at
time~$t$. In addition, in the classical model, the total energy of
the system is constant during this interval for all states of the
system. Therefore, by the generalized function rule, the relation
\begin{equation} \label{eqn:Hamiltonian-constraint-2}
 \langle \cmatrix{H}_t \rangle_t  = \langle \cmatrix{H}_{t+\Delta t}
 \rangle_{t+\Delta t}
\end{equation}
holds for any state~$\cvect{v}$. Hence, from
Eqs.~\eqref{eqn:Hamiltonian-constraint-1}
and~\eqref{eqn:Hamiltonian-constraint-2}, it follows that
\begin{equation} \label{eqn:Hamiltonian-constraint-3}
 \langle \cmatrix{H}_t \rangle_t  = \langle \cmatrix{H}_t
 \rangle_{t+\Delta t}
\end{equation}
for all~$\cvect{v}$.  But
\begin{equation}
    \langle \cmatrix{H}_t \rangle_{t+\Delta t}
        =  \langle \cmatrix{H}_t \rangle_t +
        i \langle [\cmatrix{K}_t, \cmatrix{H}_t] \rangle_t \Delta t +
        \text{O}(\Delta t^2).
\end{equation}
Therefore, the commutator~$[\cmatrix{K}_t, \cmatrix{H}_t]=0$,
which implies that there exist~$N$ mutually orthogonal
eigenvectors,~$\cvect{v}_1, \dots, \cvect{v}_N$,
which~$\cmatrix{K}_t$ and $\cmatrix{H}_t$ share in common. In
particular, the state~$\cvect{v}_j$~$(j=1,2, \dots, N)$ is an
eigenvector of~$\cmatrix{H}_t$, with some eigenvalue~$E_j$.

Now, if the system is in an eigenstate,~$\cvect{v}_j$, with
eigenvalue~$k_j$, of~$\cmatrix{K}_t$, at time~$t$, the state
evolves as
\begin{equation} \label{eqn:v-K-evolution}
    \cvect{v}(t+\Delta t) = e^{-ik_j\Delta t}\,\cvect{v}(t).
\end{equation}
Therefore, during the interval~$[t, t+ \Delta t]$, this state
remains an eigenstate of~$\cmatrix{H}_t$, and is therefore a state
of constant energy,~$E_j$, during this interval.  In addition, since
evolution only affects the overall phase of the state, the
observable degrees of freedom of the state are time-independent
during this interval.  But, by Postulate~3.4, the
state~$\cvect{v}(t)$, representing a system in a time-independent
background of definite energy,~$E_j$, whose observable degrees of
freedom are time-independent, evolves as
\begin{equation} \label{eqn:v-E-evolution}
  \cvect{v}(t+\Delta t) = e^{-iE_j\Delta t/\alpha}\,\cvect{v}(t).
\end{equation}
By comparison of Eqs.~\eqref{eqn:v-K-evolution}
and~\eqref{eqn:v-E-evolution}, we find that~$k_j = E_j/\alpha$
holds for~$j=1, 2, \dots, N$, which implies that~$\cmatrix{K}_t =
\cmatrix{H}_t/\alpha$.  Hence, in a time-independent background,
any state~$\cvect{v}$ evolves as
\begin{equation}
    \cvect{v}(t+\Delta t) =
        \exp\left(-i\cmatrix{H}_t\Delta t/\alpha \right)
        \cvect{v}(t).
\end{equation}

In order to generalize to the case of temporal evolution in a
time-dependent background, we split the interval~$[t, t+\Delta t]$
into intervals of duration~$\epsilon$, approximate the evolution
during each of these intervals assuming that the background is
time-independent, and then take the limit as~$\epsilon \rightarrow
0$:
\begin{equation}
\begin{split}
    \cmatrix{U}_t(\Delta t) &= \cmatrix{U}_{t+\Delta t -
            \epsilon}(\epsilon) \dots \cmatrix{U}_{t+\epsilon}(\epsilon)
            \cmatrix{U}_{t}(\epsilon) \\
                            &=\lim_{\epsilon \rightarrow 0}
                            \bigl\{ \cmatrix{V}_{t+\Delta t -
            \epsilon}(\epsilon) \dots \cmatrix{V}_{t+\epsilon}(\epsilon)
            \cmatrix{V}_{t}(\epsilon) \bigr\},
\end{split}
\end{equation}
which, upon expansion, yields
\begin{equation}
    \cmatrix{U}_t(\Delta t) = I - \frac{i}{\alpha}
                        \cmatrix{H}_t \Delta t + \text{O}(\Delta t^2),
\end{equation}
so that
\begin{equation}
    i\alpha\frac{d\cvect{v}(t)}{dt}  =  \cmatrix{H}_t \cvect{v}(t).
\end{equation}
The value of the constant~$\alpha$ will be determined at the end
of Sec.~\ref{sec:position-momentum-relations}.

\section{Representation of Measurements and Symmetry Transformations}
\label{sec:commutation-relations}

In this section, we shall use the AVCP to obtain~(a)~the commutation
relations~$[\cmatrix{x},\cmatrix{p}_x]=i\hbar$ and~$[\cmatrix{L}_x,
\cmatrix{L}_y]=i\hbar\cmatrix{L}_z$~(and cyclic permutations
thereof), and a restricted form of Dirac's Poisson bracket rule,
(b)~the explicit form of the displacement and rotation operators,
and (c)~the relation between the momentum and displacement
operators, and between the angular momentum and rotation operators.

From this point onwards, we shall take the infinite-dimensional
form of the abstract quantum formalism as a given.

\subsection{Position, momentum, and displacement operators}
\label{sec:position-momentum-relations}

We shall proceed in five steps.  First, by considering a particular
system~(a photon moving along the $x$-axis), we shall obtain the
commutation relationship,~$[\cmatrix{x}, \cmatrix{p}_x]=i\alpha$.
Second, we shall obtain the explicit co-ordinate representation of
the displacement operator~$\cmatrix{D}_x = \frac{1}{i}
\frac{d}{dx}$.  Third, we shall obtain the
relationship~$\cmatrix{D}_x = \cmatrix{p}_x/\alpha$ that holds
between the displacement operator,~$\cmatrix{D}_x$,
and~$\cmatrix{p}_x$, and thereby obtain the co-ordinate
representation of~$\cmatrix{p}_x$. Fourth, we shall show that the
relations obtained are generally valid. Finally, we shall make the
identification~$\alpha = \hbar$.

\subsubsection{The position--momentum commutation relationships}
\label{sec:position-momentum-relations-1}

Consider a photon moving in the $+x$-direction, where measurements
of the $x$-component of position, the $x$-component of momentum, and
the energy, are represented by the operators~$\cmatrix{x},
\cmatrix{p}_x, \cmatrix{H}$, respectively.

First, to determine the relationship of~$\cmatrix{H}$ to the
operators~$\cmatrix{x}$ and~$\cmatrix{p}_x$, we make use of the fact
that the relation~$H = cp_x$ holds for all classical
states~$(x,p_x)$ of the system, so that, from the function rule, it
follows that~$\cvect{H} = c\cvect{p}_x$.

Next, to obtain a relation between~$\cmatrix{x}$
and~$\cmatrix{p}_x$, we make use of the fact that, in the quantum
model, the expected value of~$x$ at time~$t+\delta t$ can be
calculated in two separate ways. First, from the definition
of~$\ev{x}_t$, the relation
\begin{equation}
\begin{split}
 \ev{\cmatrix{x}}_{t+\delta t} &= \bigl\langle \cmatrix{U}_t^\dag(\delta t) \,\cmatrix{x}\, \cmatrix{U}_t(\delta t) \bigr\rangle_t       \\
                    &= \left\langle \left( 1+ \frac{i\cmatrix{H}\delta t}{\alpha}\right)
                    \cmatrix{x}
                        \left( 1- \frac{i\cmatrix{H}\delta t}{\alpha}\right) \right\rangle_t  + O(\delta t^2) \\
                    &= \ev{\cmatrix{x}}_t + \frac{i}{\alpha} \delta t \ev{\cmatrix{Hx} - \cmatrix{xH}}_t      + O(\delta t^2)     \\
                    &= \ev{\cmatrix{x}}_t - \frac{ic}{ \alpha} \delta t \evb{[\cmatrix{x},\cmatrix{p}_x]}_t + O(\delta
                    t^2)
\end{split}
\end{equation}
holds for all states,~$\cvect{v}$, of the system. Second, using the
generalized function rule, it follows from the classical
relation~$x(t+\delta t) = x(t) + c\delta t + O(\delta t^2)$ that the
relation
\begin{equation}
\ev{\cmatrix{x}}_{t+\delta t} = \ev{\cmatrix{x}}_t + c\delta t +
O(\delta t^2)
\end{equation}
holds for all~$\cvect{v}$.

Equating the above two expressions for~$\ev{x}_{t+\delta t}$, we
obtain
\begin{equation}
    \cvect{v}^\dagger \left[\cvect{x}, \cvect{p}_x\right] \cvect{v} = i\alpha
\end{equation}
for all~$\cvect{v}$, which implies that
\begin{equation} \label{eqn:x-p-commutation-relation}
    \left[\cvect{x}, \cvect{p}_x \right] = i\alpha.
\end{equation}
Although a particular system has been used to obtain this
commutation relation, we shall later argue that it is generally
valid.

\subsubsection{Co-ordinate representation of the displacement
operator.}

Suppose that, in frame~$S$, the system is in state~$\psi(x)$.  The
probability density function over~$x'$ in the frame~$S'$, which is
displaced a distance~$-\epsilon$ along the $x$-axis, can be
calculated in two equivalent ways, according to whether the
transformation from frame~$S$ to~$S'$ is treated as a passive or
active transformation.  Accordingly, the probability density
function over~$x'$ can be obtained by performing measurements
of~$x'$ in frame~$S'$ upon the system in state~$\psi(x)$, or by
performing measurements of~$x$ in frame~$S$ upon the system in the
transformed state,~$\exp(-i\epsilon\cmatrix{D}_x)\psi(x)$, and
substituting~$x$ for~$x'$ in the resulting probability density
function over~$x$.

First, in frame~$S'$, let us calculate the probability density
function over~$x'$ directly.  In this frame, the
operator~$\cmatrix{x}'$ represents a measurement of~$x'$.  In the
classical model of the system, the relation
\begin{equation}
x' = x + \epsilon
\end{equation}
holds for all states~$(x,p_x)$ of the system.  Hence, by the
function rule, we obtain the operator relation
\begin{equation}
\cmatrix{x}' = \cmatrix{x} + \epsilon.
\end{equation}
Hence, an eigenstate of~$\cmatrix{x}$ with eigenvalue~$x$ is an
eigenstate of~$\cmatrix{x}'$ with eigenvalue~$x' = x + \epsilon$.
Therefore, if a measurement of~$x$ on a system in state~$\psi(x)$
yields values in the interval~$[x, x+ \Delta x]$ with
probability~$|\psi(x)|^2\Delta x$, then a measurement of~$x'$ on a
system in the same state yields values in the interval~$[x', x' +
\Delta x']$ with probability density
\begin{equation} \label{eqn:path1}
\Pr\left(x'|S', \psi(x) \right) =
\left|\psi(x'-\epsilon)\right|^2.
\end{equation}

Second, in frame~$S$, measurement~$x$ is performed on the system
in the transformed state~$\exp(-i\epsilon\cmatrix{D}_x) \psi(x)$,
so that the probability density function over~$x$ is
\begin{equation} \label{eqn:path2}
\Pr\left(x|S, \exp(-i\epsilon\cmatrix{D}_x)\psi(x) \right) =
                    \left|\exp(-i\epsilon\cmatrix{D}_x) \psi(x)\right|^2.
\end{equation}

The probability density functions over~$x'$ in
Eq.~\eqref{eqn:path1} and over~$x$ in Eq.~\eqref{eqn:path2}, must
agree under the correspondence~$x \leftrightarrow x'$.  Hence
\begin{equation}
\psi(x-\epsilon) = e^{i\phi(x)} \exp(-i\epsilon\cmatrix{D}_x)
                \psi(x),
\end{equation}
with~$\phi(x)$ being an arbitrary real-valued function of~$x$,
which is satisfied for any~$\epsilon$ if and only if~$\phi(x) = 0$
and
\begin{equation} \label{eqn:displacement-operator}
\cmatrix{D}_x = \frac{1}{i} \frac{d}{dx}.
\end{equation}

\subsubsection{The displacement--momentum operator relation.}

In a classical model, the state of a particle subject to
measurements of $x$-position and the $x$-component of momentum, is
given by~$(x_0, p_{x0})$ in some frame of reference,~$S$. Consider
the following two experiments.

In the first experiment, measurements of the $x$~components of
position and momentum of the particle are made in a reference
frame,~$S'$, that is displaced by a distance~$\epsilon$ along the
$-x$~axis, resulting in the state~$(x',p_x')$ of the particle
relative to the co-ordinates of frame~$S'$. According to the
classical model,
\begin{equation} \label{eqn:classical-displacement-relations}
    \begin{aligned}
    x' &= x_0 + \epsilon  \\
    p_x' &= p_{x0}.
    \end{aligned}
\end{equation}
In the second experiment, the particle is displaced a
distance~$\epsilon$ in the $+x$-direction, and measurements of
position and momentum are then performed in frame~$S$, giving the
state,~$(x, p_x)$, of the particle in frame~$S$ as~$(x_0 +
\epsilon, p_{x0})$.

In classical physics, for all states of the particle, the
state~$(x', p_x')$, determined by measurements in frame~$S'$ upon
the undisplaced particle, is numerically identical to the state~$(x,
p_x)$, determined by measurements in frame~$S$ upon the displaced
particle.  That is,
\begin{equation} \label{eqn:classical-active-passive-equivalence}
    (x', p_x') = (x, p_x)
\end{equation}
for all states,~$(x_0, p_{x0})$, of the particle.

Now consider a quantum model of the particle subject to measurements
of~$x$ and~$p_x$, and let the state of the particle be given
by~$\cvect{v}_0$ in frame~$S$. Consider the first experiment. From
Eqs.~\eqref{eqn:classical-displacement-relations}, it follows from
the generalized function rule that, in the quantum model of the
particle, the relations
\begin{equation} \label{eqn:displacment-av-relations1}
    \begin{aligned}
    \ev{\cmatrix{x}'} &= \cvect{v}_0^\dagger  \cvect{x} \cvect{v}_0 + \epsilon \\
    \ev{\cmatrix{p}_x'} &= \cvect{v}_0^\dagger \cvect{p}_x \cvect{v}_0,
    \end{aligned}
\end{equation}
hold for all quantum states,~$\cvect{v}_0$, of the system.

In the second experiment, the displacement of the particle is a
continuous, symmetry transformation of the system, and therefore can
be represented by a unitary transformation of the
state,~$\cvect{v}_0$, and, in particular, by the
operator~$\exp\left( -i\epsilon \cvect{D}_x \right)$,
where~$\cvect{D}_x$ is a Hermitian operator.  To first order
in~$\epsilon$, measurements of~$x$ and~$p_x$ performed on this state
have expected values
\begin{equation} \label{eqn:displacment-av-relations2}
    \begin{aligned}
    \ev{\cmatrix{x}} &= \cvect{v}_0^\dagger (1+i\epsilon \cvect{D}_x) \cvect{x}
        (1-i\epsilon \cvect{D}_x) \cvect{v}_0 \\
    \ev{\cmatrix{p}_x} &= \cvect{v}_0^\dagger (1+i\epsilon \cvect{D}_x) \cvect{p}_x
        (1-i\epsilon \cvect{D}_x) \cvect{v}_0.
    \end{aligned}
\end{equation}

From Eq.~\eqref{eqn:classical-active-passive-equivalence}, by the
generalized function rule, the average values~$\ev{x'}$
and~$\ev{p_x'}$ of Eqs.~\eqref{eqn:displacment-av-relations1} are
respectively equal to the average values~$\ev{x}$ and~$\ev{p_x}$ of
Eqs.~\eqref{eqn:displacment-av-relations2} for all~$\cvect{v}_0$.
Hence, we obtain that the relations
\begin{equation} \label{eqn:displacment-av-relations3}
    \begin{aligned}
    \cvect{v}_0^\dagger (1+i\epsilon \cvect{D}_x) \cvect{x}
        (1-i\epsilon \cvect{D}_x) \cvect{v}_0  &= \cvect{v}_0^\dagger \cvect{x}
                                             \cvect{v}_0 + \epsilon \\
    \cvect{v}_0^\dagger (1+i\epsilon \cvect{D}_x) \cvect{p}_x
        (1-i\epsilon \cvect{D}_x) \cvect{v}_0 &= \cvect{v}_0^\dagger
                                    \cvect{p}_x \cvect{v}_0,
    \end{aligned}
\end{equation}
hold for all~$\cvect{v}_0$ to first order in~$\epsilon$, which
yield the commutation relations
\begin{equation} \label{eqn:displacment-commutation-relations}
    \begin{aligned}
    \left[ \cvect{x}, \cvect{D}_x \right] &= i \\
    \left[ \cvect{p}_x, \cvect{D}_x \right] &= 0
    \end{aligned}
\end{equation}
From Eqs.~\eqref{eqn:x-p-commutation-relation}
and~\eqref{eqn:displacment-commutation-relations}, it follows that
\begin{equation}
\label{sec:displacement-minus-momentum-commutators}
    \begin{aligned}
    \left[\cvect{x}, \left(\cvect{D}_x - \cvect{p}_x/\alpha \right)\right] &= 0 \\
    \left[\cvect{D}_x, \left(\cvect{D}_x - \cvect{p}_x/\alpha \right) \right] &=
    0.
    \end{aligned}
\end{equation}

Now, in the co-ordinate representation, the
operators~$\cmatrix{x}$ and~$\cmatrix{D}_x$ are given by~$x$
and~$-i\alpha\, d/dx$, respectively, and one can readily show
that~$\{x, -i\alpha\, d/dx\}$ forms an irreducible
set~\footnote{See~\cite{Ballentine98}, Appendix~2.}. By Schur's
lemma~\footnote{See, for example, Ref.~\cite{Ballentine98},
Appendix~1.}, it therefore follows from
Eqs.~\eqref{sec:displacement-minus-momentum-commutators} that
\begin{equation} \label{eqn:displacement-momentum-relation1}
\cvect{D}_x = \frac{\cvect{p}_x}{\alpha} + \gamma \cvect{I},
\end{equation}
where~$\gamma$ is real since the operator~$\left(\cvect{D}_x -
\cvect{p}_x/\alpha \right)$ is Hermitian.  For a given
displacement,~$\epsilon$, the constant~$\gamma$ results in the
same overall shift of phase of any state,~$\cvect{v}$, of a
system, and therefore produces no physically observable effects on
the system.  Hence,~$\gamma$ can be set equal to zero without any
loss of generality, so that we obtain
\begin{equation} \label{eqn:displacement-momentum-relation2}
\cvect{D}_x = \frac{\cvect{p}_x}{\alpha}.
\end{equation}
Analogous relationships for the displacement operator
corresponding to displacements in the~$y$ and~$z$ directions can
be obtained in a similar way.

Finally, from Eqs.~\eqref{eqn:displacement-operator}
and~\eqref{eqn:displacement-momentum-relation2}, we find
\begin{equation} \label{eqn:x-momentum-operator}
\cmatrix{p}_x = \frac{\alpha}{i} \frac{d}{dx}.
\end{equation}

\subsubsection{Generality.} \label{sec:generality}

The representations of $x$- and~$p_x$-measurements have been
obtained above by considering, in the first step, a quantum model of
a particular physical system, namely a photon moving along the
$x$-axis.  In the general case of a photon moving in an arbitrary
direction, measurements of~$x, y, z$, and~$p_x, p_y, p_z$ are
sub-system measurements, and can therefore be represented in the
model of the composite system consisting of a photon, subject to
measurements chosen from a measurement set generated by a
measurement of~$\vect{r}=(x,y,z)$, by the operators~$x, y, z$
and~$-i\alpha \,\partial/\partial x, -i\alpha \,\partial/\partial y,
-i\alpha \,\partial/\partial z$, respectively.

These representations of measurements of position and momentum are
also more generally valid for other systems, as we shall explain
below.

\medskip
\paragraph{State-determined measurements.}

Suppose that, in the classical framework, a measurement of~$A$ is
performed on a system, and the outcome is determined by the state of
the system alone.  That is, in particular, the outcome is
independent of the background of the system or of any
parameters~(such as charge or rest mass) that describe intrinsic
properties of the system.  We shall then say that this measurement
is a \emph{state-determined} measurement.  For example, the outcome
of a position measurement on a particle is determined by the state
of the particle, and is independent of whether or not the particle
is in an electromagnetic field and is independent of the mass or
charge of the particle.  In general, any measurement of an
observable that is a function only of the degrees of freedom of the
state of the system is a state-determined measurement.  In contrast,
a measurement of the total energy of a system is, in general,
dependent upon not only the state of the system, but also upon the
background of the system, and is therefore not a state-determined
measurement.

Now, consider two quantum models of two different physical
systems, system~1 and system~2, in different backgrounds, with
respect to the measurement set generated by
measurements~$\mment{A}_1$ and~$\mment{A}_2$, respectively,
where~$\mment{A}_1$ and~$\mment{A}_2$ represent a measurement
of~$A$ performed on the respective systems. Suppose, further, that
the two models have the same dimension. If the measurement of~$A$
is state-determined when performed on both systems~1 and~2, then,
by the AVCP, it follows that~$\ev{\cmatrix{A}_1} =
\ev{\cmatrix{A}_2}$ holds for all states,~$\cvect{v}$, where
operators~$\cmatrix{A}_1, \cmatrix{A}_2$ represent
measurements~$\mment{A}_1, \mment{A}_2$, respectively. It follows
at once that the operators~$\cmatrix{A}_1, \cmatrix{A}_2$ are
identical.

Hence, provided that two systems admit classical models with respect
to a measurement of~$A$ that is state-determined, and admit quantum
models of the same dimension with respect to
measurements~$\mment{A}_1$ and~$\mment{A}_2$, the operators that
represent~$\mment{A}_1$ and~$\mment{A}_2$ in the respective models
must be identical.

Therefore, if state-determined measurements of~$x$ and~$p_x$ are
performed on any system, then, in a quantum model of the system
subject to measurements in the measurement set containing quantum
measurements that represent measurements of~$x$ and~$p_x$, where
these measurements yield a continuum of possible outcomes, their
representations are the same as those obtained above.  Therefore,
the commutation relations involving~$\cmatrix{x}, \cmatrix{p}_x$
and~$\cmatrix{D}_x$ are also generally valid.  Similar conclusions
hold for measurements of~$y, z$ and~$p_y, p_z$.

Therefore, in the case of a particle where the interaction energy in
the Hamiltonian is obtained from a scalar potential that is
dependent on position only, in which case the measurements of
position and momentum are state-determined, the above
representations are valid. Below, we shall consider a physically
important case where the measurement of momentum is not
state-determined.

\medskip
\paragraph{Particle in a magnetic field.}

In the case of a charged particle in a magnetic field background
described in the Hamiltonian framework, the state of the particle
is~$(\vect{x}, \dot{\vect{x}})$, but the generalized co-ordinates
are taken to be~$(\vect{x}, \vect{p})$, where~$\vect{p} =
m\dot{\vect{x}} + e\vect{A}$, where~$m$ and~$e$ are the mass and
charge of the particle, respectively, and~$\vect{A}=(A_x, A_y, A_z)$
is the vector potential.  In this case,~$\vect{p}$ depends both upon
the state of the particle and the state of the background.
Therefore, a measurement of~$\vect{p}$ is not a state-determined
measurement, and the foregoing argument cannot be used to argue that
the operators representing the measurements of the components,~$p_x,
p_y, p_z$, of~$\vect{p}$ are those derived above. Instead, we reason
as follows.

First, for a particle with state~$(\vect{x}, \dot{\vect{x}})$ in a
magnetic field, in the argument of
Sec.~\ref{sec:position-momentum-relations-1}, the commutation
relation for the $x$-component of the motion in
Eq.~\eqref{eqn:x-p-commutation-relation} becomes
\begin{equation} \label{eqn:x-x-dot-commutator}
\bigl[\cmatrix{x}, m\dot{\cmatrix{x}} \bigr] = i\alpha,
\end{equation}
Then, from~$\vect{p} = m\dot{\vect{x}} + e\vect{A}$, the sum rule
gives
\begin{equation}
\cmatrix{p}_x = m \dot{\cmatrix{x}} + eA_x(\cmatrix{x},
\cmatrix{y}, \cmatrix{z}),
\end{equation}
which, together with Eq.~\eqref{eqn:x-x-dot-commutator}, implies
that
\begin{equation} \label{eqn:canonical-commutator-in-A-field}
\left[\cmatrix{x}, \cmatrix{p}_x \right] = i\alpha,
\end{equation}
as before.

Second, we note that, in the classical framework, the
momentum~$\vect{p}$ as defined above is invariant under
displacement of the reference frame.  Therefore,
Eqs.~\eqref{eqn:displacment-commutation-relations} remain
unchanged, and, using
Eq.~\eqref{eqn:canonical-commutator-in-A-field}, we
obtain~$\cmatrix{D}_x = \cmatrix{p}_x/\alpha$.    Third, and
finally, the argument leading to the co-ordinate representation
of~$\cmatrix{D}_x$ remains unchanged since the argument only
involves measurements of position, which are state-determined
measurements. Therefore, the explicit representation
of~$\cmatrix{p}_x$ remains that given in
Eq.~\eqref{eqn:x-momentum-operator}, and similarly for the~$y$-
and~$z$-components of the motion.

\subsubsection{Identification of~$\alpha = \hbar$.}

At this point, having obtained explicit representations for position
and momentum measurements, it is possible to use the operator rules
to write down the explicit Schroedinger equation for a structureless
electron in a hydrogen atom. By solution of the equation, and by
comparing the energy levels of the electron either with those found
in Bohr's model or with those found by experiment, one can establish
that the constant~$\alpha$ is equal to~$\hbar$.

\subsubsection{Remark on applications.}

The formal rules derived above allow the quantum theoretic modeling
of a non-relativistic particle in an arbitrary classical background
consisting of gravitational and electromagnetic fields, which leads
to the non-relativistic Schroedinger equation. In the case of a
multi-particle system, the rules~(not discussed here) for dealing
with identical particles are, additionally, required.

In addition, the above rules allow the modeling of a photon without
consideration of polarization degrees of freedom~(leading to a
complex wave equation), a structureless relativistic
particle~(leading to the Klein-Gordon equation), and a relativistic
particle with internal degrees of freedom~(which, with the
appropriate auxiliary assumptions, leads to the Dirac equation).

\subsection{Angular momentum and rotation operators}

We shall proceed in three steps.  First, we shall obtain the
commutation relation~$[\cmatrix{L}_x, \cmatrix{L}_y] =
i\hbar\cmatrix{L}_z$, and cyclic permutations thereof, up to an
additive constant.  Second, we shall obtain the commutator relations
that hold between the rotation operators,~$\cmatrix{R}_x,
\cmatrix{R}_y, \cmatrix{R}_z$, and the angular momentum operators,
and shall then use these relations to determine the value of the
additive constant.  Third, we shall determine the relations that
hold between the rotation and angular momentum operators, and
indicate how the explicit representations of the angular momentum
operators and rotation operators can be determined.

\subsubsection{Components of Angular Momentum.}

Consider an experimental set-up where, in the classical model of the
set-up, measurements are performed upon a classical spin, with
magnetic moment~$\boldsymbol{\mu}$, which determine the values of
the rectilinear components of angular momentum of the system.
Suppose that the measurements of the components of angular momentum
along the $x$-, $y$-, and~$z$-directions, and the measurement of
energy, are represented by the operators~$\cmatrix{L}_x,
\cmatrix{L}_y, \cmatrix{L}_z$, and~$\cmatrix{H}$, respectively.

In particular, consider a set-up where a magnetic
field,~$\vect{B}$, is applied to the spin.  In the classical model
of this set-up, the energy associated with the interaction
is~$-\boldsymbol{\mu}\cdot\vect{B}$. Since~$\boldsymbol{\mu} = q
\vect{L}/2m$, where~$q$ and~$m$ are the charge and mass,
respectively, of the spin, and~$\vect{L}=(L_x, L_y, L_z)$ is its
angular momentum vector, the energy can be written
as~$-(q/2m)\,\vect{B}\cdot\vect{L}$. By the sum
rule~(Sec.~\ref{sec:AVCP-case2}), the quantum mechanical
Hamiltonian is given by
\begin{equation}
\cmatrix{H} = -\frac{q}{2m} (B_x\cmatrix{L}_x + B_y\cmatrix{L}_y +
B_z\cmatrix{L}_z),
\end{equation}
where~$B_x, B_y$ and~$B_z$ are the rectilinear components
of~$\vect{B}$.

The application of a magnetic field to a classical spin causes its
angular momentum vector,~$\vect{L}$, to rotate about the axis
along which the magnetic field is applied by an angle that is
proportional both to~$|\vect{B}|$ and to the duration for which
the field is applied.  Let the rotation matrix corresponding to a
rotation about axis~$a$ be denoted~$\rmatrix{R}_a(\theta)$,
where~$\theta$ is the angle of rotation. From the properties of
rotation matrices, it follows that
\begin{equation} \label{eqn:rotation-matrix-relations}
 \rmatrix{R}_x (\epsilon) \rmatrix{R}_y (\epsilon) -
    \rmatrix{R}_y (\epsilon) \rmatrix{R}_x (\epsilon) =
 \rmatrix{R}_z (\epsilon^2) - \rmatrix{I},
\end{equation}
where~$\epsilon$ is an infinitesimal angle,
and~$\rmatrix{R}_a(\epsilon)$ is an infinitesimal rotation which can
be implemented by application of a magnetic field~$\vect{B}$ along
the axis~\textit{a} for some time~$\delta t$.  Using this
relationship, it is possible to deduce the commutation relations
that hold between the quantum mechanical operators,~$\cmatrix{L}_x,
\cmatrix{L}_y, \cmatrix{L}_z$, in the following way.

The unitary evolution corresponding to the application of a
magnetic field~$\vect{B}$ to a spin for a time~$\delta t$ is
\begin{equation} \label{unitary-for-magnetic-field}
\cmatrix{U}(\delta t) = \exp \left(-\frac{i}{\hbar} \cmatrix{H}
\delta t \right).
\end{equation}
If magnetic fields of equal strength are applied for equal
times,~$\delta t$, along the $x, y,$ and~$z$-axes, respectively,
the corresponding unitary evolution is given to first order
in~$\delta t$, respectively, by
\begin{equation}
    \begin{aligned}
 \cmatrix{U}_1(\delta t) &= 1- \frac{i\epsilon}{\hbar} \cmatrix{L}_x  \\
 \cmatrix{U}_2(\delta t) &= 1- \frac{i\epsilon}{\hbar} \cmatrix{L}_y  \\
 \cmatrix{U}_3(\delta t) &= 1- \frac{i\epsilon}{\hbar} \cmatrix{L}_z.
    \end{aligned}
\end{equation}

Define~$\proj(\cvect{v})$ as the operation upon the quantum
state,~$\cvect{v}$, of a spin which returns a three-dimensional
vector,~$\ev{\vect{\cmatrix{L}}}$, with
components~$\ev{\cmatrix{L}_x}, \ev{\cmatrix{L}_y}$
and~$\ev{\cmatrix{L}_z}$.

If the application of a magnetic field,~$\vect{B} = B_z \vect{k}$,
say, to a classical spin causes a rotation of~$\vect{L}$ by
angle~$\theta$, then, by the generalized operator rule in
Eq.~\eqref{eqn:AVCP-case2-av-relation-general}, in the quantum model
of the spin, the application of the field rotates the
vector~$\ev{\vect{\cmatrix{L}}} = \proj(\cvect{v})$ by the
angle~$\theta$ about the z-axis. From
Eq.~\eqref{eqn:rotation-matrix-relations}, it therefore follows
that, for any~$\cvect{v}$,
\begin{multline}  \label{eqn:U-matrix-relations}
\proj(\cmatrix{U}_1(\epsilon) \cmatrix{U}_2(\epsilon) \cvect{v}) -
    \proj(\cmatrix{U}_2(\epsilon) \cmatrix{U}_1(\epsilon) \cvect{v} )
    = \\
    \proj(\cmatrix{U}_3(\epsilon^2) \cvect{v} ) - \proj (\cvect{v}).
\end{multline}
Using the definitions
\begin{align*}
    \cvect{v}_1 &= \cmatrix{U}_3(\epsilon^2) \cvect{v} =  \cvect{v} + \delta  \cvect{v}_1  \\
    \cvect{v}_2 &= \cmatrix{U}_1(\epsilon)\cmatrix{U}_2(\epsilon) \cvect{v}
                                         =  \cvect{v} + \delta  \cvect{v}_2     \\
    \cvect{v}_3 &= \cmatrix{U}_2(\epsilon)\cmatrix{U}_1(\epsilon) \cvect{v}
                                         =  \cvect{v} + \delta  \cvect{v}_3,     \\
\intertext{where}
    \delta  \cvect{v}_1 &= -\frac{i}{\hbar} \epsilon^2 \cmatrix{L}_z \cvect{v}    \\
    \delta  \cvect{v}_2 &=
        -\left[ -\frac{i}{\hbar} \epsilon (\cmatrix{L}_x + \cmatrix{L}_y)
                -\frac{1}{\hbar^2} \epsilon^2 \cmatrix{L}_x \cmatrix{L}_y
        \right] \cvect{v} \\
    \delta  \cvect{v}_3 &=
        -\left[ -\frac{i}{\hbar} \epsilon (\cmatrix{L}_x + \cmatrix{L}_y)
                -\frac{1}{\hbar^2} \epsilon^2 \cmatrix{L}_y \cmatrix{L}_z
        \right] \cvect{v},
\end{align*}
equation~\eqref{eqn:U-matrix-relations} becomes
\begin{equation}  \label{eqn:U-matrix-relations2}
\proj(\cvect{v} + \delta\cvect{v}_2) - \proj(\cvect{v} +
\delta\cvect{v}_3) = \proj(\cvect{v} + \delta\cvect{v}_1) -
\proj(\cvect{v}).
\end{equation}
Equating the $x$-components of this equation, we obtain
\begin{equation}
\cvect{v}^\dagger\cmatrix{L}_x (\delta \cvect{v}_2 - \delta
\cvect{v}_3) + (\delta \cvect{v}_2 - \delta \cvect{v}_3)^\dagger
\cmatrix{L}_x \cmatrix{v} = \cvect{v}^\dagger\cmatrix{L}_x \delta
\cvect{v}_1  + \delta \cvect{v}_1^\dagger \cmatrix{L}_x
\cmatrix{v},
\end{equation}
and, inserting the explicit forms of the~$\delta \cvect{v}_i$, we
obtain the commutation relation
\begin{subequations}
\begin{equation}  \label{eqn:an-operator-commutes-with-x-y-zA}
    \bigl[ \cmatrix{L}_x , \; \cmatrix{L}_z + \frac{i}{\hbar} [\cmatrix{L}_x,
        \cmatrix{L}_y ] \bigr]  =0.
\end{equation}
Equating the $y$- and $z$-components similarly, one obtains the
relations
\begin{align} \label{eqn:an-operator-commutes-with-x-y-z}
    \bigl[ \cmatrix{L}_y , \; \cmatrix{L}_z + \frac{i}{\hbar} [\cmatrix{L}_x,
        \cmatrix{L}_y ] \bigr] &=0 \\
    \bigl[ \cmatrix{L}_z , \; \cmatrix{L}_z + \frac{i}{\hbar} [\cmatrix{L}_x,
        \cmatrix{L}_y ] \bigr] &=0
        \label{eqn:an-operator-commutes-with-x-y-zC}
\end{align}
\end{subequations}
By inspection, the above commutation relations have the solution
\begin{subequations}
\begin{equation} \label{eqn:L-x-L-y-solution}
\bigl[ \cmatrix{L}_x, \cmatrix{L}_y \bigr] = i \hbar \cmatrix{L}_z
+
                                    i\gamma_1\cmatrix{I},
\end{equation}
where~$\gamma_1$ is real constant since the
operators~$i[\cmatrix{L}_x, \cmatrix{L}_y ]$ and~$\cmatrix{L}_z$ are
hermitian. We shall later show that this solution is, in fact, the
most general one.

The discussion leading to this result can be repeated to yield the
relations

\begin{align}
\bigl[ \cmatrix{L}_y, \cmatrix{L}_z \bigr] &= i \hbar \cmatrix{L}_x
                + i\gamma_2\cmatrix{I} \label{eqn:L-y-L-z-solution} \\
\bigl[ \cmatrix{L}_z, \cmatrix{L}_x \bigr] &= i\hbar\cmatrix{L}_y
        + i\gamma_3\cmatrix{I}.\label{eqn:L-z-L-x-solution}
\end{align}
\end{subequations}
In order to determine the values of $\gamma$-factors, we require
the commutation relations between the rotation
operators,~$\cmatrix{R}_x, \cmatrix{R}_y, \cmatrix{R}_z$,
and~$\cmatrix{L}_x, \cmatrix{L}_y, \cmatrix{L}_z$, which we shall
now derive.

\subsubsection{Rotation--angular momentum commutation relations.}

Let an infinitesimal clockwise rotation of a frame of reference by
angle~$\epsilon$ about the~$z$-axis be represented by unitary
transformation~$\exp(-i\epsilon \cmatrix{R}_z)$,
where~$\cmatrix{R}_z$ is Hermitian.

Now consider a set-up where measurements of~$L_x, L_y$, and~$L_z$
are performed on a system in the original and in the transformed
frame of reference.  In the classical model of this situation, the
outcomes of the measurements performed in the original~(unprimed)
and transformed~(primed) frames are, to first order in~$\epsilon$,
related by
\begin{equation} \label{eqn:classical-rotation-connections}
\begin{pmatrix} L_x' \\ L_y' \\ L_z'\end{pmatrix} =
                                \begin{pmatrix}
                                    1 & -\epsilon & 0 \\
                                    \epsilon & 1 & 0 \\
                                    0 &   0 &   1
                                \end{pmatrix}
                                \begin{pmatrix}
                                    L_x \\ L_y \\ L_z
                                \end{pmatrix}.
\end{equation}

By the generalized operator rule in
Eq.~\eqref{eqn:AVCP-case2-av-relation-general}, it follows that, in
the quantum model of the situation, the relation
\begin{equation} \label{eqn:av-rotation-connections}
        \begin{pmatrix} \ev{\cmatrix{L}_x'} \\ \ev{\cmatrix{L}_y'} \\ \ev{\cmatrix{L}_z'}
        \end{pmatrix}
        =
        \begin{pmatrix}
            1 & -\epsilon & 0 \\
            \epsilon & 1 & 0 \\
            0 &   0 &   1
        \end{pmatrix}
        \begin{pmatrix}
        \ev{\cmatrix{L}_x} \\ \ev{\cmatrix{L}_y} \\ \ev{\cmatrix{L}_z}
        \end{pmatrix}
\end{equation}
holds for all states,~$\cvect{v}$, of the system.  Using the
relation
\begin{equation}
\ev{\cmatrix{L}_x'} = \cvect{v}^\dagger (1+ i\epsilon
\cmatrix{R}_z)\cmatrix{L}_x (1-  i\epsilon \cmatrix{R}_z) \cvect{v}
+ O(\epsilon^2),
\end{equation}
we find from Eq.~\eqref{eqn:av-rotation-connections} that
\begin{equation}
\ev{\cmatrix{L}_x} + i\epsilon \ev{[\cmatrix{R}_z, \cmatrix{L}_x]}
=
            \ev{\cmatrix{L}_x} - \epsilon \ev{\cmatrix{L}_y}
\end{equation}
for all~$\cvect{v}$, which implies that
\begin{subequations}
\begin{equation} \label{eqn:R-z-L-x-commutator}
\bigl[\cmatrix{R}_z, \cmatrix{L}_x \bigr] = i\cmatrix{L}_y.
\end{equation}
Proceeding similarly for~$\ev{\cmatrix{L}_y'}$
and~$\ev{\cmatrix{L}_z'}$, we obtain
\begin{align}
\left[\cmatrix{R}_z, \cmatrix{L}_y\right] &= -i\cmatrix{L}_x \label{eqn:R-z-L-y-commutator} \\
\left[\cmatrix{R}_z,\cmatrix{L}_z\right] &= 0.
                                \label{eqn:R-z-L-z-commutator}
\end{align}
\end{subequations}

The commutation relations for~$\cmatrix{R}_x$ and~$\cmatrix{R}_y$
can be obtained by parallel arguments.

\subsubsection{Angular momentum commutation relations.}

Left-multiplying Eq.~\eqref{eqn:L-z-L-x-solution}
by~$\cmatrix{R}_z$, and applying
Eqs.~\eqref{eqn:R-z-L-x-commutator}--\eqref{eqn:R-z-L-z-commutator},
we obtain
\begin{equation}
\bigl[\cmatrix{L}_z, \cmatrix{L}_x \bigr] \cmatrix{R}_z - i
\bigl[\cmatrix{L}_y, \cmatrix{L}_z \bigr] = \left(i\hbar
\cmatrix{L}_y + i\gamma_3 \cmatrix{I}\right)\cmatrix{R}_z + \hbar
\cmatrix{L}_x.
\end{equation}
Using the Eq.~\eqref{eqn:L-z-L-x-solution} right-multiplied
by~$\cmatrix{R}_z$, this implies that
\begin{subequations}
\begin{equation}
\bigl[\cmatrix{L}_y, \cmatrix{L}_z \bigr] = i\hbar \cmatrix{L}_x.
\end{equation}
Parallel arguments applied to Eqs.~\eqref{eqn:L-x-L-y-solution}
and~\eqref{eqn:L-y-L-z-solution} using the commutation relations
between~$\cmatrix{R}_x, \cmatrix{R}_y$ and~$\cmatrix{L}_x,
\cmatrix{L}_y, \cmatrix{L}_z$ yield
\begin{align}
\bigl[ \cmatrix{L}_z, \cmatrix{L}_x \bigr] &= i\hbar\cmatrix{L}_y \\
\bigl[ \cmatrix{L}_x, \cmatrix{L}_y \bigr] &= i \hbar
                                                    \cmatrix{L}_z.
\end{align}
\end{subequations}

\subsubsection{Explicit form of angular momentum operators}

From the classical relation~$L^2 = L_x^2 + L_y^2 + L_z^2$, it
follows from the sum rule that~$\cmatrix{L}^2 =\cmatrix{L}_x^2
+\cmatrix{L}_y^2+\cmatrix{L}_z^2$.  Using this relation and the
above commutation relations for~$\cmatrix{L}_x, \cmatrix{L}_y$
and~$\cmatrix{L}_z$, explicit representations of these operators for
finite~$N$ can be obtained and the irreducibility of the
representations of~$\cmatrix{L}_x, \cmatrix{L}_y, \cmatrix{L}_z$ can
be shown~\footnote{See~\cite{Group-Theory-in-Physics}~(Ch.~10,
Sec.~3), for example.}. Therefore, by Schur's lemma, the solution
given in Eq.~\eqref{eqn:L-x-L-y-solution} is the most general
solution of
Eqs.~\eqref{eqn:an-operator-commutes-with-x-y-zA}--\eqref{eqn:an-operator-commutes-with-x-y-zC},
and similarly for the solutions given in
Eqs.~\eqref{eqn:L-y-L-z-solution} and~\eqref{eqn:L-z-L-x-solution}.

Although the representations of~$\cmatrix{L}_x, \cmatrix{L}_y$
and~$\cmatrix{L}_z$ have been obtained by considering a particular
physical system, they are generally valid on account of the
argument given in Sec.~\ref{sec:generality}. Therefore the
commutation relations for~$\cmatrix{L}_x, \cmatrix{L}_y$
and~$\cmatrix{L}_z$ are generally valid.

\subsubsection{Rotation--angular momentum relations,
and explicit form of the rotation operators.}

Using the commutation relationships for~$\cmatrix{L}_x,
\cmatrix{L}_y$ and~$\cmatrix{L}_z$, it follows from
Eqs.~\eqref{eqn:R-z-L-x-commutator}--\eqref{eqn:R-z-L-z-commutator}
that~$(\hbar \cmatrix{R}_z - \cmatrix{L}_z)$ commutes
with~$\cmatrix{L}_x, \cmatrix{L}_y$, and~$\cmatrix{L}_z$.
Since~$\{\cmatrix{L}_x, \cmatrix{L}_y, \cmatrix{L}_z\}$ is an
irreducible set, it follows from Schur's lemma that
\begin{equation}
\hbar \cmatrix{R}_z - \cmatrix{L}_z = \gamma \cmatrix{I},
\end{equation}
where~$\gamma$ is a real-valued constant since~$\cmatrix{R}_z$
and~$\cmatrix{L}_z$ are Hermitian.  For any given~$\epsilon$, a
non-zero value of~$\gamma$ results in the same overall change of
phase for all states transformed
by~$\exp(-i\epsilon\cmatrix{R}_z)$, and so cannot give rise to
observable consequences. Hence,~$\gamma$ can be set to zero
without loss of generality. Therefore,~$\cmatrix{R}_z =
\cmatrix{L}_z/\hbar$, and, similarly,~$\cmatrix{R}_x =
\cmatrix{L}_x/\hbar$ and~$\cmatrix{R}_y = \cmatrix{L}_y/\hbar$.

Using the explicit representations of~$\cmatrix{L}_x,
\cmatrix{L}_y, \cmatrix{L}_z$ for any given dimension~$N$, the
explicit representation of the rotation operators follows at once
from these rotation--angular momentum relations. The explicit
co-ordinate representations of the rotation operators in the
infinite-dimensional case can also be determined by an argument
similar to that used earlier to determine the explicit form of the
displacement operators.

\subsection{Commutators and Poisson brackets}
\label{sec:PB-analogy}

In this section, we shall obtain a relation between the Poisson
Bracket,~$\{A, B\}$, and the commutator~$[\cmatrix{A},
\cmatrix{B}]$, where~$A$ and~$B$ are the classical observables of
a physical system describable in the classical Hamiltonian
framework, and~$\cmatrix{A}, \cmatrix{B}$ are the operators that
represent measurements of these observables. Dirac's Poisson
Bracket rule asserts the relation
\begin{equation} \label{eqn:Dirac-PB}
[\cmatrix{A}, \cmatrix{B}] = i\hbar \widehat{\{A, B\}},
\end{equation}
where~$\widehat{\{A,B\}}$ is the operator that represents a
measurement of~$\{A, B\}$.  Below, we shall derive this relation
using the AVCP in the case where~$B$ is the Hamiltonian.

Consider the Hamiltonian model of a system with state~$(q_1,
\dots, q_N; p_1, \dots, p_N)$ where~$N\ge 1$. The temporal rate of
change of the function~$F(q_1, \dots, q_N; p_1, \dots, p_N)$ is
given in terms of the Hamiltonian,~$H(q_1, \dots, q_N; p_1, \dots,
p_N)$, by
\begin{equation} \label{eqn:F-dot-PB}
\begin{split}
\dot{F} &=  \{F,H\} \\
        & = \sum_{i=1}^{N} \left\{ \frac{\partial F}{\partial q_i}
            \frac{\partial H}{\partial p_i} -  \frac{\partial H}{\partial q_i}
            \frac{\partial F}{\partial p_i} \right\}.
\end{split}
\end{equation}
Consider the quantum model of the system with state~$\cvect{v}$,
where the measurements of the~$q_i$ and the~$p_i$ are represented by
operators~$\cvect{q}_i$ and~$\cvect{p}_i$, respectively. If~$H$ is
simple, then, by the AVCP, a measurement of~$H$ can be represented
by the operator~$\cmatrix{H}$; otherwise, according to the AVCP, it
is not possible to describe the temporal evolution of the system in
the quantum model. If both of the functions~$\dot{F}$ and~$\{F, H\}$
are simple, then, by the AVCP, they are represented by the
operators~$\widehat{\dot{F}}$ and~$\widehat{\{F,H\}}$, respectively,
and from Eq.~\eqref{eqn:F-dot-PB}, by the generalized function rule,
the relation
\begin{equation} \label{eqn:PB1}
\evb{\widehat{\dot{F}}}_t =  \evb{\widehat{\{F,H\}}}_t,
\end{equation}
holds for all~$\cvect{v}$.

Now, in the classical model, the function~$\dot{F}$ is defined,
for all states, as
\begin{equation} \label{eqn:classical-F-dot}
\dot{F} =  \lim_{\Delta t \rightarrow 0} \left\{ \frac{F(t+\Delta
t) - F(t)}{\Delta t} \right\}.
\end{equation}
If~$F(t)$ and~$F(t+\Delta t)$ are both simple, then, according to
this definition,~$\dot{F}(t)$ is also simple, and, using the
generalized sum rule~(regarding the measurement of~$F(t+\Delta t)$
as the one being implemented in terms of measurements of~$F(t)$
and~$\dot{F}(t)$), we obtain the relation
\begin{equation} \label{eqn:PB2}
\evb{\widehat{\dot{F}}}_t =\lim_{\Delta t \rightarrow 0}
\frac{1}{\Delta t}
                    \left\{ \ev{\cmatrix{F}}_{t+\Delta t} - \ev{\cmatrix{F}}_t
                    \right\},
\end{equation}
which holds for all~$\cvect{v}$, with the operator~$\cmatrix{F}$
representing a measurement of~$F$.

If the functions~$\{F, H \}$ and~$F(t)$ are both simple, then,
since~$F(t+\Delta t) = F(t) + \{F, H \} \Delta t$, it follows
that~$F(t+\Delta t)$ is also simple. In that case, both
Eqs.~\eqref{eqn:PB1} and~\eqref{eqn:PB2} hold for all~$\cvect{v}$.
Equating these two expressions for~$\ev{\widehat{\dot{F}}}_t$, we
obtain the relation
\begin{equation} \label{eqn:PB-average-relation}
\begin{split}
    \evb{\widehat{\{F,H\}}}_t &=\lim_{\Delta t \rightarrow 0} \frac{1}{\Delta t}
                     \bigg\{ \left\langle \left( 1+ \frac{i}{\hbar} \cvect{H}\Delta t\right)
                        \cvect{F}
                     \left( 1- \frac{i}{\hbar} \cvect{H}\Delta t \right)
                     \right\rangle_t \\
                            &\quad\quad\quad\quad\quad\quad - \ev{\cmatrix{F}}_t  \bigg\} \\
                &= i\hbar^{-1} \left\langle [\cvect{H},\cvect{F}]
                \right\rangle_t,
\end{split}
\end{equation}
which holds for all~$\cvect{v}$, so that
\begin{equation} \label{eqn:PB-analogy}
    i\hbar \widehat{\{F,H\}}  =  [\cvect{F}, \cvect{H}].
\end{equation}
Hence, we obtain Eq.~\eqref{eqn:Dirac-PB} in the special case
where~$\cmatrix{B} = \cmatrix{H}$, subject to the condition that
the functions~$A, B$ and~$\{A,B\}$ are simple.

Using this relationship, we can readily evaluate useful commutation
relationships.  For example, for a photon with state~$(x,p_x)$,
setting~$F = x$, and~$H = cp_x$~(the Hamiltonian for a photon moving
along the $+x$-direction), we find~$\{F,H\} = c$.  Hence, since the
functions~$F, H$, and~$\{F,H\}$ are simple,
Eq.~\eqref{eqn:PB-analogy} immediately gives~$[\cvect{x},
\cvect{p}_x] = i\hbar$.

If one or more of the functions~$F, H$, and~$\{F,H\}$ is not
simple, then Eq.~\eqref{eqn:PB-analogy} does not follow from the
above argument.
To take a specific example, suppose that, for a system with
state~$(x,p_x)$, where~$[\cmatrix{x}, \cmatrix{p}_x]=i\hbar$, we
choose~$F=x^3$ and~$H=\gamma p_x^3$, where~$\gamma$ is a constant.
We can then apply the function rule to obtain the corresponding
operators~$\cmatrix{F} =\cmatrix{x}^3$ and~$\cmatrix{H} =
\gamma\cmatrix{p}_x^3$,  and use these to find
\begin{equation} \label{eqn:pb-example-1}
[\cmatrix{F}, \cmatrix{H}] = 3i\gamma\hbar \left(\cmatrix{x}^2
\cmatrix{p}_x^2 + \cmatrix{x} \cmatrix{p}_x^2 \cmatrix{x} +
\cmatrix{p}_x^2 \cmatrix{x}^2\right).
\end{equation}
However, the function~$\{F,H\} = 9\gamma x^2p_x^2$ is not simple,
which implies that the AVCP cannot be used to write down an operator
which represents a measurement of~$\{F,H\}$.   If we were
nonetheless to apply the Hermitization rule in
Eq.~\eqref{eqn:hermitization-rule} to a measurement of~$\{F,H\}$~(in
spite of the inconsistencies to which we have shown this would lead)
we would obtain
\begin{equation} \label{eqn:pb-example-2}
i\hbar \widehat{\{F, H\}} = \frac{9i\gamma\hbar}{2}
\left(\cmatrix{x}^2\cmatrix{p}_x^2 + \cmatrix{p}_x^2 \cmatrix{x}^2
\right),
\end{equation}
but this differs from~$[\cmatrix{F}, \cmatrix{H}]$ by the
constant~$2\gamma\hbar^3$.  Since the expected value
of~$\gamma\hbar\cmatrix{x}^2\cmatrix{p}_x^2$ is itself of
order~$\gamma\hbar^3$, there is no guarantee that the difference
between Eqs.~\eqref{eqn:pb-example-1} and~\eqref{eqn:pb-example-2}
will be negligible.

The question of whether Eq.~\eqref{eqn:Dirac-PB} holds in the more
general case where the function~$B$ cannot be treated as the
classical Hamiltonian of the system is not discussed here.

\section{Arbitrariness in the functions~$f(\var_i)$
                                        and~$\tilde{f}(\var_i)$}
\label{sec:arbitrariness}

In Paper~I, we found that the functions~$f(\var_i)$
and~$\tilde{f}(\var_i)$~(not to be confused the function~$f$ that
forms part of the AVCP) are~$\pm \cos(a\var_i +b)$ and~$\pm
\sin(a\var_i +b)$, respectively, where~$a, b \in \numberfield{R}$
and~$a\neq 0$. Having obtained the temporal evolution operator and
commutation relationships, we are now able to show that the
choices made in Paper~I of the positive signs for~$f$
and~$\tilde{f}$ and of~$a=1$ and~$b=0$ do not involve a loss of
generality.

Let us first consider the case where positive signs are chosen for
both~$f$ and~$\tilde{f}$.  With constants~$a$ and~$b$ left in
place, the derivation given in Paper~I is altered as follows.
First, with~$Q_{a|i} = \cos(a\var_i +b)$ and~$Q_{b|i} =
\sin(a\var_i +b)$, and defining~$\tilde{\var}_i = a\var_i +b$, we
obtain
\begin{equation}
\label{eqn:defn-of-lvect-Q-2-general}
\begin{split}
\lvect{Q}   &= (\sqrt{P_1} Q_{a|1}, \sqrt{P_1} Q_{b|1}, , \dots, \sqrt{P_N} Q_{b|N}) \\
            &= (\sqrt{P_1} \cos\tilde{\var}_1, \sqrt{P_1} \sin\tilde{\var}_1,
            \dots, \sqrt{P_N} \sin\tilde{\var}_N),
\end{split}
\end{equation}
where the~$P_i$ are the probabilities of the observed outcomes of
the measurement,~$\mment{A}$, with respect to which~$\lvect{Q}$ is
written.

Second, the invariance condition~(Postulate~3.2) requires that
there is no change in the probabilities of the observable outcomes
of measurement~\mment{A} performed on an evolved state if an
arbitrary constant~$\var_0 \in \numberfield{R}$ is added to each
of the~$\var_i$ in the initial state.  Hence, in terms of
the~$\tilde{\var}_i$, an arbitrary constant~$\tilde{\var}_0$ may
be added to each of the~$\tilde{\var}_i$.  Therefore, from the
argument given in Paper~I, the state can, with respect to
measurement~$\mment{A}$, be represented as the complex vector
\begin{equation}
\cvect{v} = \begin{pmatrix}
            \sqrt{P_1} e^{i\tilde{\var}_1} \\
            \sqrt{P_2} e^{i\tilde{\var}_2} \\
            \vdots                \\
            \sqrt{P_N} e^{i\tilde{\var}_N} \\
            \end{pmatrix},
\end{equation}
with physical transformations being represented as unitary or
antiunitary transformations as found previously.

Third, Postulate~5 implies that~$\tilde{\var}_{ij} =
\tilde{\var}^{(1)}_i + \tilde{\var}^{(2)}_j$, which implies that
the composite systems rule remains~$\cvect{v} = \cvect{v}^{(1)}
\otimes \cvect{v}^{(2)}$, where~$\cvect{v}^{(1)}, \cvect{v}^{(2)}$
are the states of the sub-systems, and~$\cvect{v}$ is the state of
the composite system.

Fourth, by inspection of the argument used above to find the
temporal evolution operator, one finds that~$\cmatrix{U}_t(\Delta t)
= \exp(-ia\cmatrix{H}\Delta t/\alpha)$.  In addition, the foregoing
arguments lead to~$\cmatrix{p}_x = -i(a/\alpha)\, d/dx$. If one
writes down the Schroedinger equation implied by these relations,
one obtains~$\alpha/a = \hbar$, which implies
that~$\cmatrix{U}_t(\Delta t) = \exp(-i\cmatrix{H}\Delta t/\hbar)$
as before.

In summary, irrespective of the values of~$a, b$, one obtains the
same abstract quantum formalism.  Furthermore, although the
constants~$a, b$ appears as a connection between~$\var_i$
and~$Q_{a|i}$, we shall now show that they are unimportant insofar
as observable predictions of the formalism, namely the prediction of
the values and probabilities of the observable outcomes of
measurements, are concerned.  To do so, we shall consider an
experiment where a system is prepared using measurement~$\mment{A}$
and, after undergoing a physical transformation~(either active or
passive), is subject to measurement~$\mment{A}'$.

We shall first establish that the explicit form of a measurement
operator that represents any given measurement is independent of~$a$
and~$b$.  First, consider the example of the operators~$\cmatrix{x}$
and~$\cmatrix{p}_x$.  The explicit form of these operators is
determined by (a)~the commutation relation~$[\cmatrix{x},
\cmatrix{p}_x]=i\hbar$, (b)~the measurement--transformation
relation~$\cmatrix{D}_x = \cmatrix{p}_x/\hbar$, and (c)~the
relation~$\cmatrix{D}_x = -i\,d/dx$. In this case, one finds by
inspection of the above derivation that these relations are all
independent of~$a$ and~$b$.  More generally, the explicit form of
measurement operators is determined by (a)~commutation relationships
between measurement operators, (b)~the relation between measurement
operators and transformation operators, and (c)~the explicit form of
transformation operators.  First, since the Dirac Poisson bracket
rule derived in Sec.~\ref{sec:PB-analogy} is independent of~$a$
and~$b$, the commutation relations between measurement operators are
independent of~$a$ and~$b$.  Second, the measurement--transformation
relations and the explicit form of the transformation relations are
directly obtained via the AVCP from the classical relation between
the outcomes of measurements performed in the original and
transformed frames of reference, and are therefore independent
of~$a$ and~$b$~(as illustrated by the above arguments leading
to~$\cmatrix{D}_x = \cmatrix{p}_x/\hbar$ and $\cmatrix{D}_x =
-i\,d/dx$).  Therefore, in general, measurement operators are
independent of~$a$ and~$b$.

Next, we consider the operators that represent physical
transformations.  First consider active transformations.  The
general temporal evolution operator,~$\cmatrix{U}_t(\Delta t) =
\exp(-i\cmatrix{H}_t\Delta t/\hbar)$ is not explicitly dependent
upon~$a$ and~$b$.  The operator~$\cmatrix{H}_t$ is a function of
the measurement operators~(such as~$\cmatrix{x}, \cmatrix{p}_x$
for a particle moving along the $x$-axis) relevant to the system,
and is directly obtained~(via the AVCP) from the classical
relation between the classical Hamiltonian,~$H$, and the classical
observables that determine the state of the system, and is
therefore also independent of~$a$ and~$b$. Since the measurement
operators are themselves independent of~$a$ and~$b$, it follows
that the operator representing any given temporal evolution is
independent of~$a$ and~$b$.

Second, the explicit form of operators representing passive
transformations~(such as displacement or rotation of a reference
frame) can be obtained by directly transposing the relevant
classical relations~(which connect the co-ordinates in the
original and transformed frames) into the quantum framework via
the~AVCP~(as illustrated by the derivation of the co-ordinate
representation of~$\cmatrix{D}_x$ above), and can be readily seen
to therefore also be independent of~$a$ and~$b$.  Hence, a unitary
or antiunitary operator that represents any given physical
transformation is independent of~$a$ and~$b$.

Now consider the above-mentioned experiment where
measurement~$\mment{A}$ is performed and, say, outcome~$1$ is
obtained.  With respect to measurement~$\mment{A}$, the resulting
state is~$\cvect{v} = (1, 0 \dots, 0)^{\text{\textsf{T}}}$ up to an
irrelevant overall phase.  Suppose the system now undergoes a
physical transformation~(either active or passive), and a second
measurement,~$\mment{A}$, is performed. Since the physical
transformation is represented by a unitary or antiunitary operator
that is independent of~$a$ and~$b$ as noted above, the outcome
probabilities of the second measurement are unaffected by the value
of~$a$ and~$b$.  This example includes the general case where the
second measurement is measurement~$\mment{A}'$ since the latter can,
by Postulate~1.2~(see Paper~I), be represented by an arrangement
consisting of measurement~$\mment{A}$ immediately preceded and
followed by suitable temporal evolution of the system. Therefore, in
the most general experiment we are considering, the outcome
probabilities of the observable outcomes are also independent of~$a$
and~$b$.

Finally, we note that, since the explicit form of the measurement
operators is independent of~$a$ and~$b$, their eigenvalues are
also independent of~$a$ and~$b$. Therefore, in summary, we find
that both the probabilities and the values of the observable
outcomes of measurements in the above general experiment are
independent of~$a$ and~$b$. Therefore, the values of~$a$ and~$b$
can, without loss of generality, conveniently be chosen to
be~$a=1$ and~$b=0$.

If one chooses the signs of~$f$ and~$\tilde{f}$ not to be both
positive, then this is equivalent to choosing positive signs
for~$f$ and~$\tilde{f}$ but changing the values of~$a$ and~$b$ to
some other values,~$a'$ and~$b'$, respectively. Specifically, if
one chooses the signs~$(+,-)$, then~$a' = -a$ and~$b' = -b+\pi$;
if~$(-,+)$, then~$a' = -a$ and~$b' = -b$; and, if~$(-,-)$,
then~$a' = a$ and~$b' = b+\pi$.  But we have already shown that
the choice of the constants~$a$ and~$b$ is unimportant, and hence
the change in their values is unimportant. Therefore, the signs
can, without loss of generality, both be chosen to be positive.

\section{Discussion}

The derivation presented in this paper has shown that, using a
clearly-motivated physical principle, it is possible to derive the
explicit form of the temporal evolution operator given the
postulates of Paper~I, and to derive the formal rules of quantum
theory in a systematic manner from appropriately chosen relations
known to hold in classical physics. The derivation provides several
physical insights into the formal rules.

The first insight is that the classical description of a
measurement~(such as `a measurement of~$x^2$') leaves open the
possibility of more than one implementation, and that, when
modeled in the quantum framework, these implementations are, in
general, not equivalent.

Second, it is possible to impose a simple average-value condition
that must be satisfied by an operator that can be said to represent
a classical implementation of a measurement.  This condition implies
that many implementations cannot be represented by an operator, and
can therefore be eliminated from consideration. That is, one finds
that there are implementations which, although acceptable in the
classical framework, cannot be represented by operators in the
quantum framework without violating a very mild average-value
condition.

Third, in the case of an implementation that satisfies the
average-value condition, the operator that represents the
implementation is uniquely determined by the average-value condition
provided that the function,~$f$, that describes the implementation,
is simple. One also finds that those implementations that satisfy
the average-value condition are are represented by the same
operator, so that it is possible to represent a measurement of~$f$
by a unique operator. If~$f$ is not simple, then it does not appear
to be possible to apply the average-value condition, even in a
weakened form, without inconsistencies arising.

Fourth,  we have found that the AVCP is incompatible with the
assumption that every classically-described measurement on a system
is represented by a quantum measurement in the quantum model of the
system.  For example, the AVCP implies that a measurement of~$AB$
does not have an operator representation if~$[\cmatrix{A},
\cmatrix{B}]\neq 0$.

The fifth insight rests on the fact that, rather surprisingly, the
AVCP enables formal rules of each of the four types described in the
Introduction~(operator rules, commutation relations, transformation
operators, and measurement--transformation relations) to be obtained
in a uniform manner. Consequently, one can see that the difference
between these types of rules depends simply upon whether the
classical relations that one is taking over into the quantum
framework are relations between measurements performed at the same
time~(leading to the operator rules), at different times~(leading to
commutation relations for measurement operators), or in different
frames of reference~(leading to measurement--transformation
relations and to the explicit forms of transformation operators). In
short, from the perspective provided by the derivation, the
commutation relation~$[\cmatrix{x}, \cmatrix{p}_x] = i\hbar$ is no
more elusive in its origin than
the operator relation~$\cmatrix{H} = \cmatrix{p}_x^2/2m$.

Sixth, the derivation provides a clearer physical foundation to many
particular formal rules that are commonly used in quantum theory.
For example, the commutation relation~$[\cmatrix{L}_x,
\cmatrix{L}_y]=i\hbar \cmatrix{L}_z$ is ordinarily derived in the
infinite-dimensional quantum formalism for a particle~(by
transposing the classical relation~$\cmatrix{L}_z = xp_y - yp_x$,
and cyclic permutations thereof, into the quantum framework using
the operator rules), and is then assumed, without further
justification, to also hold in the finite-dimensional case. Here, we
have obtained this commutation relation directly for finite- and
infinite-dimensional quantum systems, and have done so in a manner
that makes clear its connection with the properties of rotations.
Similarly, a restricted form of Dirac's Poisson bracket rule has
been derived in a systematic manner using the AVCP without making
use of abstract analogies.

Finally, we remark that the general notion of average-value
correspondence is already familiar in elementary quantum mechanics
through Ehrenfest's theorem~\cite{Ehrenfest-average-values}, which
shows that the motion of a particle modeled in the quantum framework
is, on average, approximated by the behavior of the particle when
described classically.  However, the possibility that such a
correspondence might serve as the basis for a constructive principle
that allows the formal rules of quantum theory to be determined by
appropriately-chosen classical relations does not appear to have
been widely
explored~\footnote{%
Refs.~\cite{vNeumann55, Groenewold-Principles-QM, Bohm51} mention
the general idea of average-value correspondence in their
discussion of the operator rules of quantum theory.  For example,
Groenewold~\cite{Groenewold-Principles-QM}~(Eqs.~(1.32)--(1.34))
remarks that the sum rule is equivalent to a condition on the
expectations of the respective operators, but the idea is not
formulated in a manner that is sufficiently systematic to derive
the operator rules, and no attempt is made to derive the any of
the other types of formal rule~(such as the commutation relations)
using average-value correspondence.  Bohm~\cite{Bohm51} clearly
articulates the idea that average-value correspondence can be used
as a constraint on quantum theory, and uses it to determine
particular instances of the function rule~(Secs.~9.5--9.21) and to
determine the Hamiltonian operator that represents a
non-relativistic particle~(Secs.~9.24--9.26). However, the idea is
not
systematically formulated and applied beyond these special cases.}.  %
It has been shown here that it is possible to formulate the notion
of average-value correspondence in the form of an exact physical
principle which, roughly speaking, allows the logic of Ehrenfest's
argument to be reversed, enabling the often physically obscure
formal rules of quantum theory to be derived in a systematic manner
from familiar relations known to hold in classical physics.

\begin{acknowledgments}

I am indebted to Steve Gull and Mike Payne for their constant
support and invaluable comments, and to Tetsuo Amaya and Matthew
Donald for discussions and invaluable comments.
\end{acknowledgments}

\end{document}